%% file: main.tex
\documentclass[sigconf]{acmart}

\AtBeginDocument{%
  \providecommand\BibTeX{{%
    \normalfont B\kern-0.5em{\scshape i\kern-0.25em b}\kern-0.8em\TeX}}}

\setcopyright{acmcopyright}

\copyrightyear{2019}
\acmYear{2019}
\acmConference[SC '19]{The International Conference for High Performance Computing, Networking, Storage, and Analysis}{November 17--22, 2019}{Denver, CO, USA}
\acmBooktitle{The International Conference for High Performance Computing, Networking, Storage, and Analysis (SC '19), November 17--22, 2019, Denver, CO, USA}
\acmPrice{15.00}
\acmDOI{10.1145/3295500.3356162}
\acmISBN{978-1-4503-6229-0/19/11}

\usepackage{booktabs} 

\usepackage{amsmath,amssymb,amsfonts}
\usepackage{graphicx}
\usepackage{textcomp}
\usepackage{graphicx}
\usepackage{latexsym}
\usepackage[]{algorithm2e}
\usepackage{algpseudocode}
\usepackage{courier}
\usepackage{listings}
\usepackage{xcolor}
\usepackage{amssymb}
\usepackage{amsthm}
\usepackage{amsmath}
\usepackage{multicol}
\usepackage{lipsum}
\usepackage{comment}
\usepackage{amsmath}
\usepackage{subfig}
\usepackage[export]{adjustbox}
\usepackage{listings}
\usepackage[inline]{enumitem}
\usepackage[all]{nowidow}
\usepackage{array}
\usepackage{tikz}
\usepackage[skip=6pt]{caption}

\newcommand{\bhline}{\noalign{\hrule height 1pt}}
\newcolumntype{?}{!{\vrule width 1pt}}
\newcolumntype{^}{!{\vrule width 1.5pt}}

\definecolor{myblue}{RGB}{91,155,213}
\definecolor{mydark}{RGB}{0,0,0}

\begin{document}

\title[Software Systolic Array Execution Model (SSAM)]{A Versatile Software Systolic Execution Model for GPU Memory-Bound Kernels}

\author{Peng Chen}
\orcid{1234-5678-9012}
\affiliation{%
  \institution{Tokyo Institute of Technology}
  \institution{AIST-Tokyo Tech Real World Big-Data Computation Open Innovation Laboratory, National Institute of Advanced Industrial Science and Technology}
}
\email{chen.p.aa@m.titech.ac.jp}

\author{Mohamed Wahib}
\orcid{1234-5678-9012}
\affiliation{%
  \institution{AIST-Tokyo Tech Real World Big-Data Computation Open Innovation Laboratory, National Institute of Advanced Industrial Science and Technology}
}
\email{mohamed.attia@aist.go.jp}

\author{Shinichiro Takizawa}
\orcid{1234-5678-9012}
\affiliation{%
  \institution{AIST-Tokyo Tech Real World Big-Data Computation Open Innovation Laboratory, National Institute of Advanced Industrial Science and Technology}
}
\email{shinichiro.takizawa@aist.go.jp}

\author{Ryousei Takano}
\orcid{1234-5678-9012}
\affiliation{%
  \institution{National Institute of Advanced Industrial Science and Technology}
}
\email{takano-ryousei@aist.go.jp}

\author{Satoshi Matsuoka}
\orcid{1234-5678-9012}
\affiliation{%
  \institution{Tokyo Institute of Technology}
  \institution{RIKEN Center for Computational Science, Hyogo, Japan}  
}
\email{matsu@acm.org}

\renewcommand{\shortauthors}{Chen, P. and Wahib, M., et al.}

\begin{abstract}
This paper proposes a versatile high-performance execution model, inspired by systolic arrays, for memory-bound regular kernels running on CUDA-enabled GPUs. We formulate a systolic model that shifts partial sums by CUDA warp primitives for the computation. We also employ register files as a cache resource in order to operate the entire model efficiently. We demonstrate the effectiveness and versatility of the proposed model for a wide variety of stencil kernels that appear commonly in HPC, and also convolution kernels (increasingly important in deep learning workloads). Our algorithm outperforms the top reported state-of-the-art stencil implementations, including implementations with sophisticated temporal and spatial blocking techniques, on the two latest Nvidia architectures: Tesla V100 and P100. For 2D convolution of general filter sizes and shapes, our algorithm is on average {
2.5$\times$ faster} than Nvidia's NPP on V100 and P100 GPUs.
\end{abstract}

\ccsdesc[500]{Computer systems organization~Systolic arrays}
\ccsdesc[500]{Computer systems organization~Multicore architectures}

\keywords{Systolic Array, GPU, CUDA, Convolution, Stencil}

\maketitle

\input{body-conf}

\bibliographystyle{ACM-Reference-Format}
\bibliography{main}

\end{document}

%% file: body-conf.tex
\section{Introduction}\label{sec:introduction}

GPU accelerators have been increasingly adopted to meet the exponentially growing computational requirements in various fields, such as scientific simulations and machine learning. Those increasing computational requirements are pushing for the trend of building GPU-accelerated supercomputers made up of dense nodes that include several GPUs (e.g. ORNL Summit has six GPUs/node, and LLNL's Sierra and TokyoTech's Tsubame 3.0 both have four GPUs/node). Hence it is imperative that codes running on those systems be high in performance and scale vertically (i.e. on a single node), as well as horizontally (i.e. on the entire system). In this paper, we focus on the single node performance of one of the commonly occurring computational motifs in HPC (and occasionally deep learning): memory-bound regular computation on a structured grid~\cite{Asanovic:2009:VPC:1562764.1562783}. 

Memory-bound kernels that have a regular pattern of computation are particularly challenging, since they appear to be simple, yet they require very complex data reuse schemes to effectively utilize the memory hierarchy. Typically, advanced GPU implementations for memory-bound kernels on structured grids rely on the optimized use of fast on-chip scratchpad memory: the programmer uses this user-managed scratchpad memory for reducing the global memory access. Indeed, there exists a plethora of work proposing variations and combinations of the three locality schemes that rely on scratchpad memory: spatial blocking, temporal blocking, and a wavefront pipeline (the reader can find some of the notable work at~\cite{maruyama2014optimizing,Micikevicius:2009:FDC:1513895.1513905,5645463,6569833,Yuan:2017:TS:3126908.3126920,DBLP:journals/siamsc/MalasHLSWK15}). Those complex locality schemes enabled strides in performance improvements. However, they essentially moved the bottleneck from the global memory to the faster, yet smaller, scratchpad. The objective of this work is to yet again move the bottleneck from the scratchpad to a faster resource: register files.

This paper proposes a versatile systolic execution model for improving the performance of memory-bound kernels with regular access patterns. A wide class of kernels and applications can benefit from this versatile model: convolution, stencils, scan/reduction operators~\cite{merrill2015cub}, Summed Area Tables~\cite{chen2018efficient}, \dots etc. The systolic model is based on the transfer and accumulation of partial results in thread-private registers. Additionally, we employ the register files as a cache to avoid using the scratchpad altogether.

To accumulate and transfer partial sums using thread-private registers, in a SIMT fashion, different groups of threads (known as warps in CUDA) operate over different input points, with some data redundancy that we introduce to account for the halo layers. Different threads in a warp compute the partial sums, before moving the partial sums to the downstream neighbor thread to be accumulated. To transfer the partial sums, we rely on the warp shuffle primitives that provide low-latency register exchange within a warp~\cite{TESLA-V100}. 


To match the high throughput of shuffling the partial sums, we fully utilize the registers for caching the computed partial sums. Accordingly, our model can perform structured grid memory-bound computations at low latency and high throughput. As a result, we can decrease the dependency on scratchpad or cache memory and thus improve the application's performance by avoiding the scratchpad and cache bottleneck. To avoid overemphasis on intra-warp communication, it is necessary to clarify that we do not limit the use of scratchpad for \emph{inter-warp communication}.

A systolic array model~\cite{mccanny1984optimised} is typically a well-structured two-dimensional mesh of Processing Elements (PEs) and provides extremely high TOPS (TeraOps/Second) and high TOPS/Watt~\cite{jouppi2017datacenter}. Recently, systolic arrays have been successfully used to speed up deep learning workloads, e.g. Google Tensor Processing Unit (TPU)~\cite{jouppi2017datacenter} and Nvidia Tensor Cores.
Inspired by the mechanism of hard-wired systolic arrays, we propose \emph{a versatile execution model} to improve the performance of regular memory-bound kernels by moving the memory access bottleneck from the scratchpad to registers. Our model can be viewed as Software Systolic Array Execution Model (SSAM).

{
Kernels that can be mapped to SSAM should have a regular memory access pattern. However, there are several challenges that should be addressed in order to implement an efficient software systolic model on the top of partial sums accumulation/exchange, and register caching.
First, the algorithms must be expressed in the way the systolic array can process. 
Second, the total number of threads executed together in a working unit (i.e. CUDA warp) is relatively small\footnotemark\footnotetext{\emph{WarpSize} is equal to 32 on all Nvidia GPU generations}. Hence, this enforces a limit on the systolic array size and parallelism. We have to rely on Instruction Level Parallelism (ILP) and in-thread data reuse to provide enough concurrency in each node (i.e. thread) in the systolic array.
Finally, the shuffle primitives work only for a single warp: a redundancy method for halo layers, along with a redundancy analysis, is necessary. 
}
The contributions in this paper are as follows:
{
    \setlength{\leftmargini}{10 pt}
    \begin{itemize} 
    \item A formulation, design, and implementation of a model (called SSAM),
    inspired by systolic arrays, for efficiently computing memory-bound kernels with regular access on GPUs.
    \item A detailed analysis of the data reuse and redundancy schemes to quantify the efficiency and limitations of SSAM.
    \item Evaluation of the proposed model for a wide variety of iterative 2D/3D stencils and 2D general convolution on Tesla P100/V100 GPUs. Our model outperforms the top reported state-of-the-art implementations, including implementations with sophisticated temporal and spatial blocking techniques. 
    {
    SSAM-based convolution is over 2.5$\times$ faster than Nvidia's NPP library, and up to 1.5$\times$ faster than the ArrayFire library.}
    \end{itemize}
}
{
The rest of this paper is organized as follows. 
Section~\ref{sec:bkground} discusses the background, i.e., CUDA, convolution, and stencils.
In Section~\ref{sec:ssam} we propose the formulation of SSAM.
In Section~\ref{sec:ssam-imp} we present the implementation of various algorithms in SSAM.
Section~\ref{sec:performance-model} describes our performance model.
In Section~\ref{sec:Evaluation} we report the evaluated performance of convolution and stencils on the latest Nvidia GPUs.
Section~\ref{sec:discussion} discusses the GPU architectures from the perspective of operation dependencies in SSAM.
In Section~\ref{sec:Related Work}, we review the related work.
Finally, Section~\ref{sec:Conclusion} concludes.
}

\section{Background}\label{sec:bkground}
We briefly introduce the CUDA's concepts and programming model (more details about CUDA can be found in~\cite{cudaToolkit}):  
{
    \begin{enumerate*}[label=(\roman*)]
    \item {\bf{CUDA Memory Hierarchy}}. 
    GPUs support different memory types: global, local, texture, constant, shared and register files~\cite{cudaToolkit}. 
    Global memory is the largest off-chip memory with the highest R/W (Read/Write) latency. Shared memory is a fast on-chip scratchpad memory limited in scope to CUDA thread blocks. 
    \item{\bf{Register Cache Vs. Shared Memory Cache.}}
        Register cache is an approach in which a single warp builds a virtual cache layer on top of register files for low-latency data R/W~\cite{ben2016fast}. 
        As Table~\ref{tbl:Shared Memory and Register Files} shows, on the latest GPUs, the register memory per SM is 256KB ({{65536*4B}}) and is more than 2.7$\times$ larger than shared memory. 
    \item {\bf{Intra-Warp Communication By Shuffle.}}
    Shuffle is an intra-warp communication mechanism for a CUDA-enabled GPU. It allows the exchange of data between threads directly within a single warp without using shared memory or global memory. 
    \item {\bf{Register Spilling.}}
    Registers are a limited resource in GPUs. If the required number of registers per thread is too high, the compiler may spill to local memory~\cite{micikevicius2011local}. However, there are methods to reduce the register pressure as in~\cite{RegDem, Rawat:2018:AIR:3291656.3291718}.
    \end{enumerate*}
}

\subsection{Convolution}
\label{sec:Computing 2D Convolution}
Mathematically, convolution combines two linear functions to form a third one in order to measure the correlation of overlap between functions. The canonical form of 2D convolution is
\begin{equation*}\small \begin{split}
(f \ast w)(x,y) = \sum_{t=c}^d \sum_{s=a}^b f(x - s, y - t){\cdot}w(s, t)
\label{equ:2d conv}
\end{split}
\end{equation*}
Where $\ast$ and $\cdot$ are convolution and multiplication operators, respectively. f(x, y) denotes a 2D matrix (or image), while w is a filter of size (M, N) applied to the matrix, where M=b-a+1, N=d-c+1.
Hereafter, we assume that the size of the matrix is (W, H) where W is the width of the matrix and H is its height. 

Note that throughout the paper, we use W, H, M, N as defined in this section.

\label{sec:background}
    \begin{table}[t]  
          \caption{Shared Memory and Register Files on GPUs}  
          \small
          \centering
          \begin{tabular}{ ? c | c | c | c ?}
          \bhline
          {\bf{Tesla GPU}} & {\bf{Shared Memory/SM}} & {\bf{32-bit registers/SM}} & {\bf{SMs}}\\
          \bhline
          K40  & 16/32/48 KB & 65536 & 15\\
          M40  & 96 KB & 65536 & 24\\ 
          P100 & 64 KB & 65536 & 56\\
          V100 & up to 96 KB   & 65536 & 80\\
          \bhline
          \end{tabular}  
          \label{tbl:Shared Memory and Register Files} 
      \end{table}  
\begin{figure*}[t]
\subfloat[2D stencil.] {
   \includegraphics[width=0.13\textwidth]{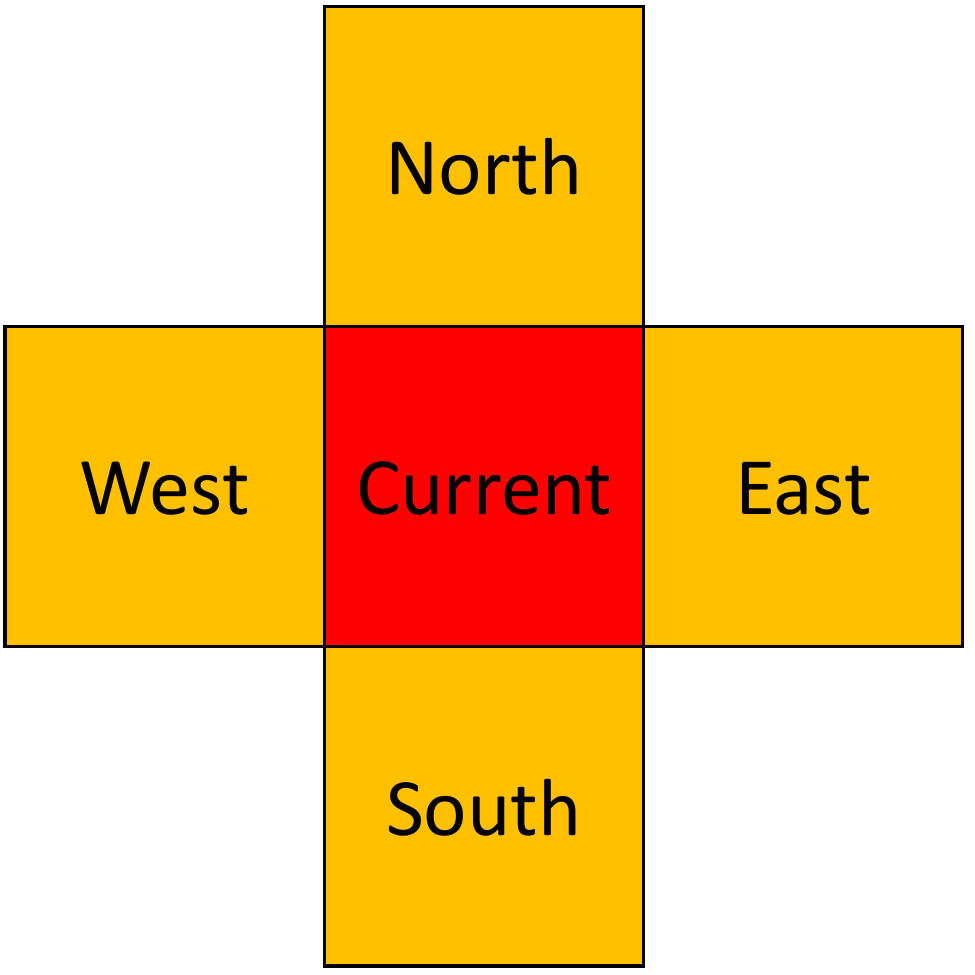}
   \label{fig:2dstencil}
 }
\hspace{0.00\textwidth}
\subfloat[3D stencil.] {
   \includegraphics[width=0.13\textwidth]{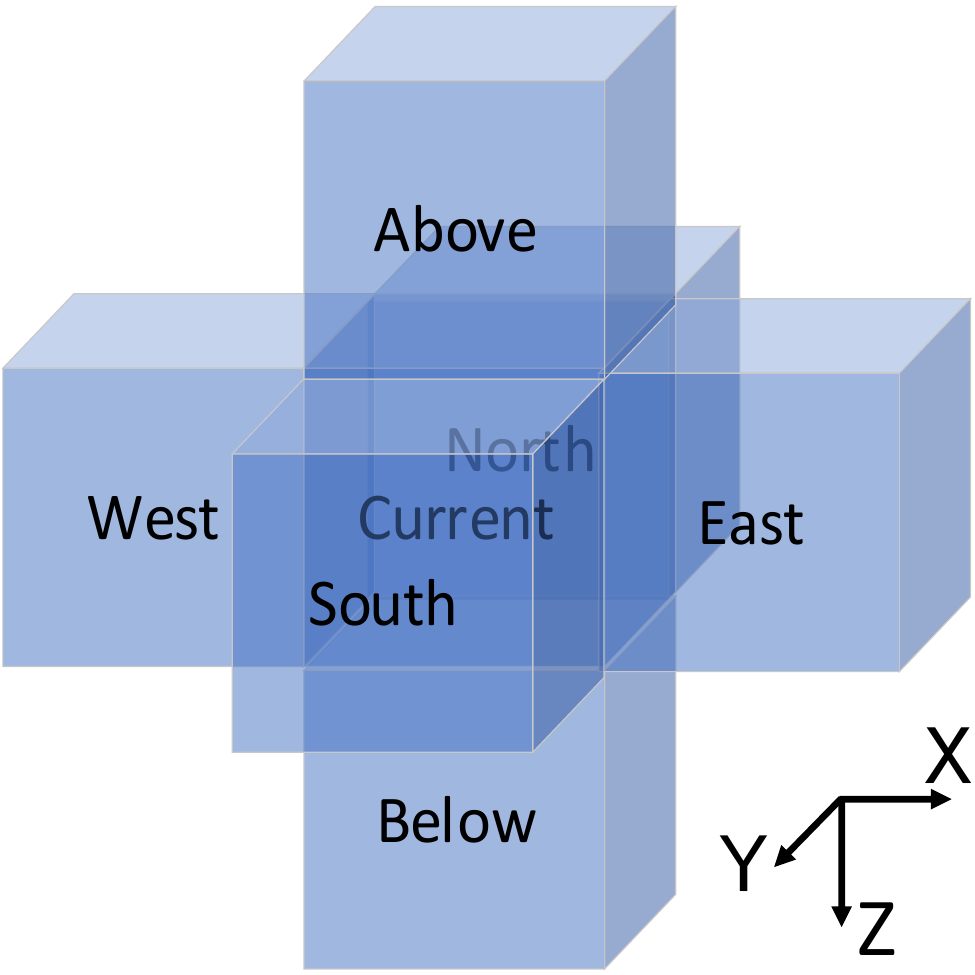}
   \label{fig:3dstencil}
 }
\hspace{0.0315\textwidth}
\subfloat[Hardware Systolic Array.] {
   \includegraphics[width=0.19\textwidth]{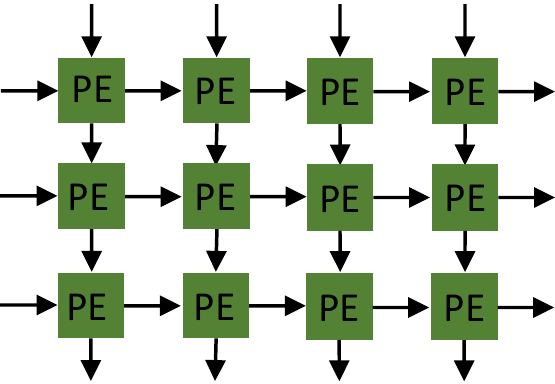}
   \label{fig:hwsystolicarray}
 }
\hspace{0.00\textwidth}
\subfloat[Software Systolic Array.] {
   \includegraphics[width=0.22\textwidth]{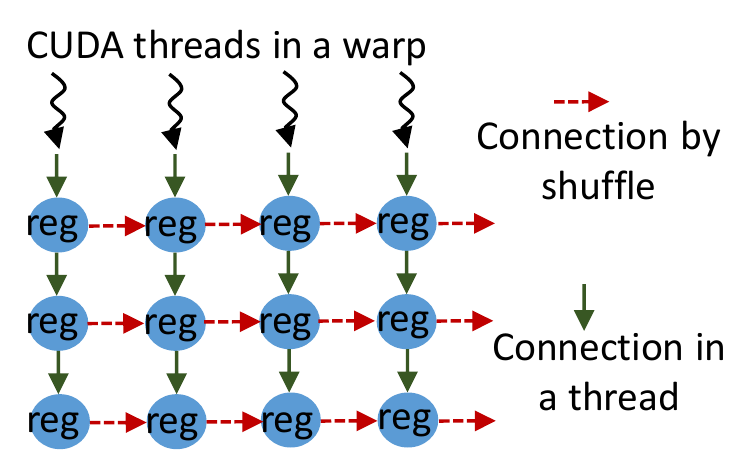}
   \label{fig:ssai}
 }
\hspace{0.0312\textwidth}
\subfloat[KS-scan for 8 elements~\cite{kogge1973parallel}.] {
   \includegraphics[width=0.20\textwidth]{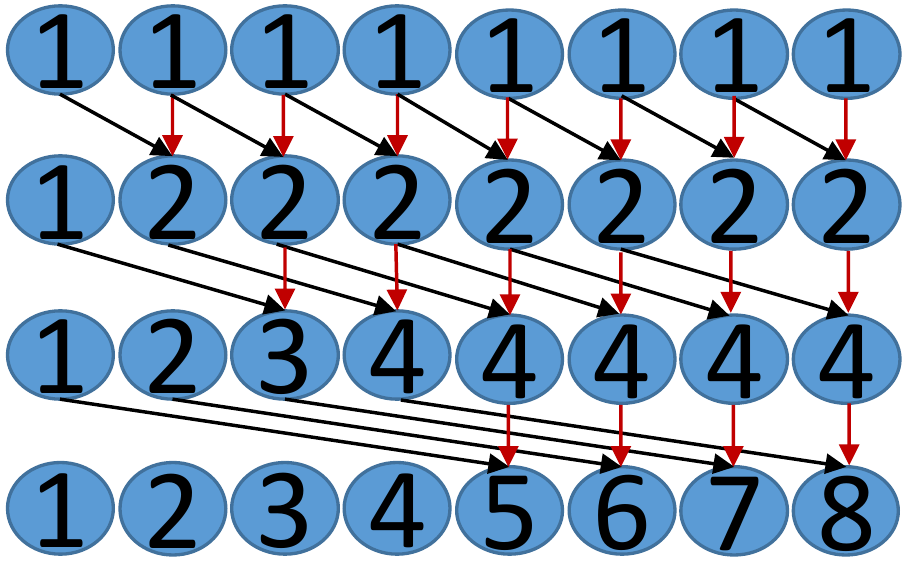}
   \label{fig:ks-scan}
 }
\caption{(a) 2D 5-point stencil example. (b) 3D 7-point stencil example. (c) Hardware 2D systolic array structure. PE is a processing element. (d) SSAM on CUDA (2D problem illustration: "reg" is a register). In the vertical direction, registers are in the same thread. In the horizontal direction, registers are exchanged by the shuffle instruction. (e) Eight elements KS-scan~\cite{kogge1973parallel}. 
Arrows represent the dependency of our model in Equation~\ref{equ:model J}.
} 
\end{figure*}
\subsection{Stencils}\label{sec:setncils}
Iterative stencil computations are fundamental for scientific applications in many domains~\cite{milne1953numerical, talay1994numerical}.
Figure~\ref{fig:2dstencil} shows a typical first-order 2D diffusion 5-point stencil (also known as 2D Jacobi stencil). 
Mathematically, the 2D diffusion stencil is defined as
\begin{equation*}\small
\begin{split}
s_{i+1}(x, y) &= s_{i}(x-1, y){\cdot}West+s_{i}(x, y-1){\cdot}North \\
        &~+s_{i}(x, y)\cdot Current+s_{i}(x, y+1){\cdot}South+s_{i}(x+1, y){\cdot}East
\label{equ:2d stencil}
\end{split}
\end{equation*}
where (x, y) is a relative position, $s_{i}$ is a cell's value at the $i^{th}$ iteration, and $\cdot$ is the multiplication operator. West, North, Current, South, and East denote the stencil coefficients. 

Regarding 3D stencils, taking the 7-point diffusion stencil as an example (Figure~\ref{fig:3dstencil}), two vertical points (namely Above and Below) are added to the 2D 5-point stencil. The stencil is computed by adding $s_{i}(x, y, z-1){\cdot}Above$ and $s_{i}(x, y, z-1){\cdot}Below$, where (x, y, z) is the cell position. 

\section{SSAM: Software Systolic Array Model}\label{sec:ssam}
We propose an execution model, named Software Systolic Array Execution Model (SSAM), that enhances the performance of regular memory-bound kernels on GPUs. 
Our work is motivated by the recent revival of systolic arrays and similar architectures such as Tensor cores of Nvidia GPUs, Intel TBB~\cite{reinders2007intel} data flow graphs, and wavefront accelerators (mainly FPGAs~\cite{kilts2007advanced} and ASICs~\cite{bhatnagar2002advanced}). Tensor cores are specially optimized for matrix multiplication, however, SSAM is a general model that is applicable to a wide range of memory-bound algorithms. 
\subsection{Systolic Arrays}
A systolic array~\cite{kung1982systolic} is a structured computing model composed of many interconnected, yet independent, Processing Elements (PEs) as shown in Figure~\ref{fig:hwsystolicarray}. In parallel, all PEs compute the partial results for a specified function, store results locally, and then send them to neighbor PEs for the next cycle. The computational and data storage behavior of each PE can be formulated as 
\begin{equation}\small \begin{split}
s{\leftarrow}ctrl(r{\otimes}x){\oplus} s 
\label{equ:partial result}
\end{split} \end{equation}
where \emph{s} is a partial result held by each PE, \emph{r} is an external coefficient (e.g. for convolution \emph{r} is the weights), and \emph{x} is an input value. Both $\otimes$ and $\oplus$ are basic arithmetic operations, and $ctrl(E)$ is a control function having the value of \emph{0} or \emph{E}, i.e. the output of $ctrl(E)$ is \emph{0} or \emph{E}.
\subsection{Hardware Systolic Array}
In the past decades, systolic arrays have been well researched. The authors in \cite{kung1979let, quinton1983systematic, yeh1984systolic, li1985design, cheng2005novel} have proposed well-structured systolic arrays architectures. Ideally, a systolic array is qualified for a specific computing purpose: e.g., matrix multiplication~\cite{kumar1991synthesizing}, convolution~\cite{kung1981hardware}, DCT~\cite{lee1990simple} \dots etc. A hardware systolic array is a computing pipeline network with physical interconnects between PEs. As Figure~\ref{fig:hwsystolicarray} shows, all of the PEs are connected with their neighbors. The PEs simultaneously compute, store, and transfer their partial results to downstream neighbors. 

\subsection{SSAM: Software Systolic Array Model}
{
SSAM is built as an execution model to express a variety of algorithms, with regular access pattern, in a systolic array fashion. In other words, it simulates the mechanism of hardware systolic arrays in CUDA architecture. Our target in this paper is to improve the performance of the memory-bound kernels with regular memory access pattern. However, SSAM, in general, is not limited to memory-bound kernels and could be extended to compute bound kernels, such as GEMM. In this section, we present SSAM's formulation and discuss motivating examples.
}
Three core techniques contribute to the SSAM model:
{
    {
        \setlength{\leftmargini}{16 pt}
        \begin{enumerate}[label=(\roman*)]
        \item Efficient in-register computation via register cache;
        \item Fast intra-warp communication via shuffle instructions;
        \item Parallel accumulation of partial sums.
        \end{enumerate}
    }
}
We build a software systolic array on the top of a virtual register cache that is used for efficient data access and reuse. In addition, the shuffle instructions are used for low latency communication between threads in a warp. As Figure~\ref{fig:ssai} shows, each register performs the same function as a PE in Figure~\ref{fig:hwsystolicarray}. It is important to iterate that the registers (PEs) in the vertical direction belong to the same CUDA thread, while the registers in the horizontal direction belong to different threads.

The similarity between systolic array PEs and our registers can be listed as follows: 
\begin{enumerate*}[label=(\roman*)]
\item the ability to execute any arithmetic operations by oneself
\item the ability to store partial result by oneself
\item the ability to pass the partial result to neighbors (direct register access in the vertical direction, and using shuffle in the horizontal direction), 
\item most importantly, all of the operations are performed simultaneously.
\end{enumerate*}

It is worth mentioning that the use of in-register computing and partial sums have been proposed in specific implementations of individual algorithms (as will be discussed in the related work in Section~\ref{sec:Related Work}). On the other hand, the model proposed in this paper is a robust and versatile model that can be leveraged by a wide variety of problems with different patterns of data dependency, as will be demonstrated in the following sections. 

\subsection{Expressing Algorithms in SSAM}
In order to express algorithms in SSAM, we build an algorithmic formulation that extends the prior work in the literature~\cite{moldovan1987advis}. From the perspective of a CUDA warp, an algorithm can be formulated as a four-tuple 
\begin{equation}\small \label{equ:model J}\begin{split}
\emph{J}=(O, D, X, Y)
\end{split}\end{equation}
where
\begin{itemize}[noitemsep]
\item \emph{O} : computing operations of \emph{J}
\item \emph{D} : dependencies of \emph{J}
\item \emph{X} : input variables of \emph{J}
\item \emph{Y} : output variables of \emph{J}
\end{itemize}
In the model \emph{J}, the register cache is used for storing \emph{X} and \emph{Y} (more details about register cache will be discussed in Section~\ref{sec:regster cache}). In contrast to hardware systolic arrays, SSAM can perform computing operations \emph{O} like Equation~\ref{equ:partial result}, i.e. all arithmetic and intrinsic operations supported by CUDA. In addition, the graph representing the dependencies \emph{D} can take any shape and is not limited to a structured mesh. This is due to the fact that the shuffle operation allows arbitrary threads in a warp to exchange values held in registers. It is worth mentioning that the partial results are updated as specified in Equation~\ref{equ:partial result}, and are passed according to the dependency graph \emph{D}.
{
There are several methods to extract the dependency graph \emph{D}. We briefly introduce one of those methods in Section~\ref{sec:automation-ssam} : using the polyhedral model~\cite{Griebl98codegeneration} to represent D.
}
\subsection{Motivating Example 1: 1D Convolution} \label{sec:1-d-conv}
We use 1D convolution as a motivating example to illustrate how an algorithm can be expressed in SSAM. Given an input array $A=[a_0,\cdot\cdot\cdot,a_{n-1}]$ and filter $F=[f_0, f_1, f_2]$, the output of convolution is an array $B=[b_0,\cdot\cdot\cdot,b_{n-1}]$, such that $b_i=\sum_{z=i-1}^{i+1}{a_z}\cdot f_z$. We use registers to cache the input data, with one thread computing one element. In SSAM, we map the 1D convolution to a warp (Equation~\ref{equ:model J}) such that the input $X=[a_k,\cdot\cdot\cdot,a_{k+WarpSize-1}]$ and output $Y=[b_k,\cdot\cdot\cdot,b_{k+WarpSize-1}]$, where $0\leq k{\leq}n-WarpSize$. For the computation \emph{O} (in Equation~\ref{equ:partial result}) where r$\in$F, $\otimes$ and $\oplus$ are multiplication and addition operations, respectively. The dependency graph $D$ can be represented as in Figure~\ref{shitf_partial_sum}, 
where the $ctrl()\equiv 1$ is fixed. 
\subsection{Motivating Example 2: Scan Operator}
{\bf{Scan operator}} is the sums of prefixes (running total) of an input sequence~\cite{kogge1973parallel
}. Even though we do not evaluate Scan operator in this paper, we choose Scan operator as an example of mapping algorithms in SSAM to demonstrate its versatility. Given an input array $[a_0,\cdot\cdot\cdot,a_{n-1}]$, the output of Scan is an array $[b_0,\cdot\cdot\cdot,b_{n-1}]$, such that $b_i=\sum_0^i{a_i}$. For simplicity, we use registers to cache the input data where each thread holds one element. In our model, we map the Scan operation to a warp in SSAM such that the input $X=[a_k,\cdot\cdot\cdot,a_{k+WarpSize-1}]$, and output $Y=[b_k,\cdot\cdot\cdot,b_{k+WarpSize-1}]$, where $0\leq k \leq n-WarpSize$. As to the computation operations \emph{O}, in each computing stage (Equation~\ref{equ:partial result}), $r\equiv 1$, $\otimes$ is a multiplication and $\oplus$ is an addition operation. Since we use the Kogge-Stone Scan algorithm~\cite{kogge1973parallel}, the dependency graph \emph{D} is similar to Figure~\ref{fig:ks-scan}, namely the arrows in Figure~\ref{fig:ks-scan} can be expressed as \emph{ctrl()} in Equation~\ref{equ:partial result}. 

One-dimensional scan operator is simple to some extent. Furthermore, the complex case of {\bf{two-dimensional scan}}, known as Summed Area Tables (SAT), is proven to benefit of a systolic model optimization similar to SSAM, more details can be found in our another work~\cite{chen2018efficient}.

\section{Implementation of Algorithms in SSAM}
\label{sec:ssam-imp}
Detailed techniques required in implementing the SSAM model are illustrated in this section, such as coalescing global memory accesses, caching data by register files, computing and transferring partial sums, and overlapped blocking. Using 2D convolution as a motivating example, we discuss the implementation of SSAM. Additionally, we mention the special considerations given to stencil operators when necessary.

\subsection{Mapping 2D Convolution to SSAM}
\begin{figure*}[t]
\centering
\subfloat[Register Cache.] {
   \includegraphics[width=0.40\textwidth]{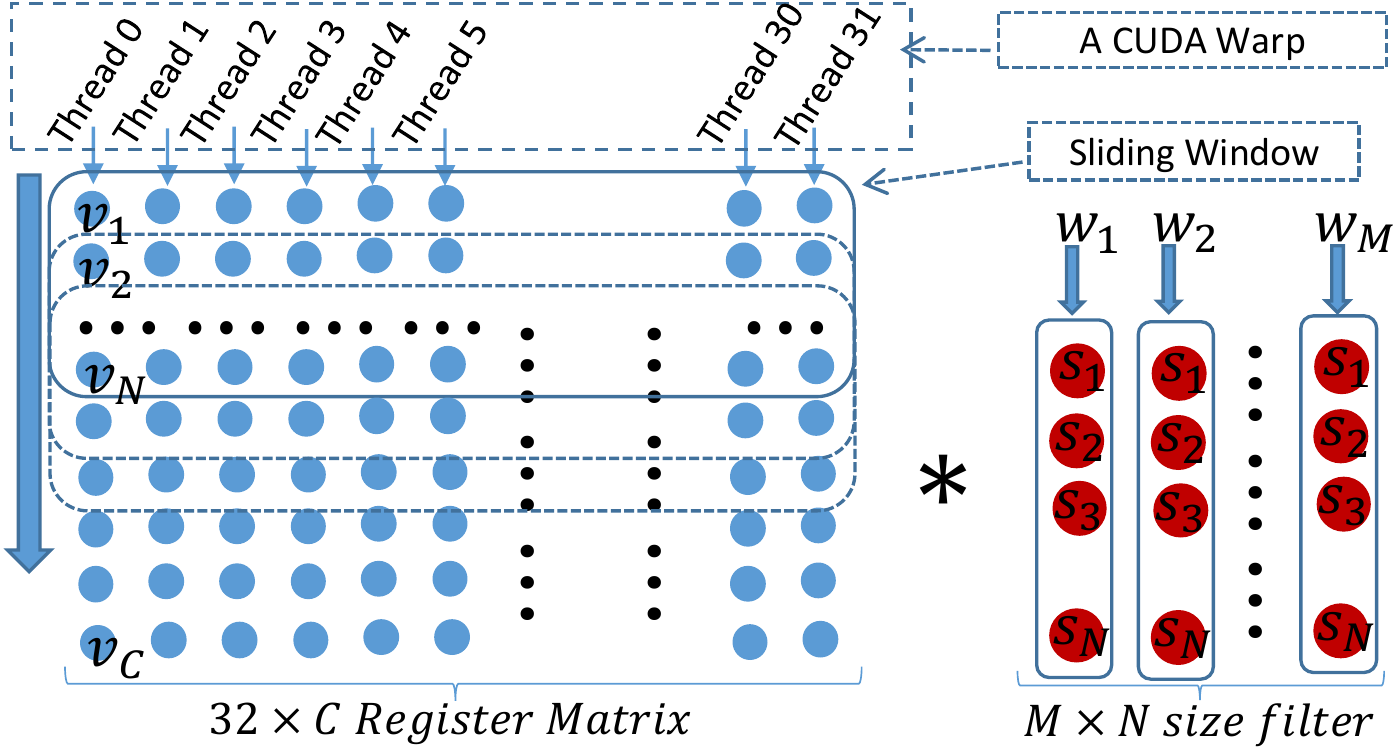}
   \label{register_cache_shared_mem_cache}
 }
 \hspace{0.025\textwidth}
\subfloat[Computing Partial Sums.] {
   \includegraphics[width=0.185\textwidth]{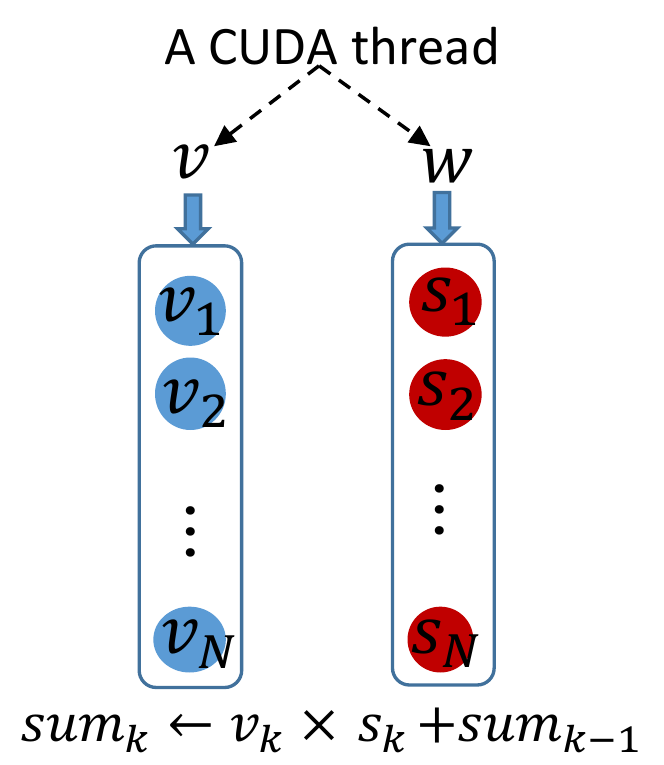}
   \label{parallel_inner_product}
 }
  \hspace{0.015\textwidth}
 \subfloat[Transfer Partial Sums.] {
   \includegraphics[width=0.33\textwidth]{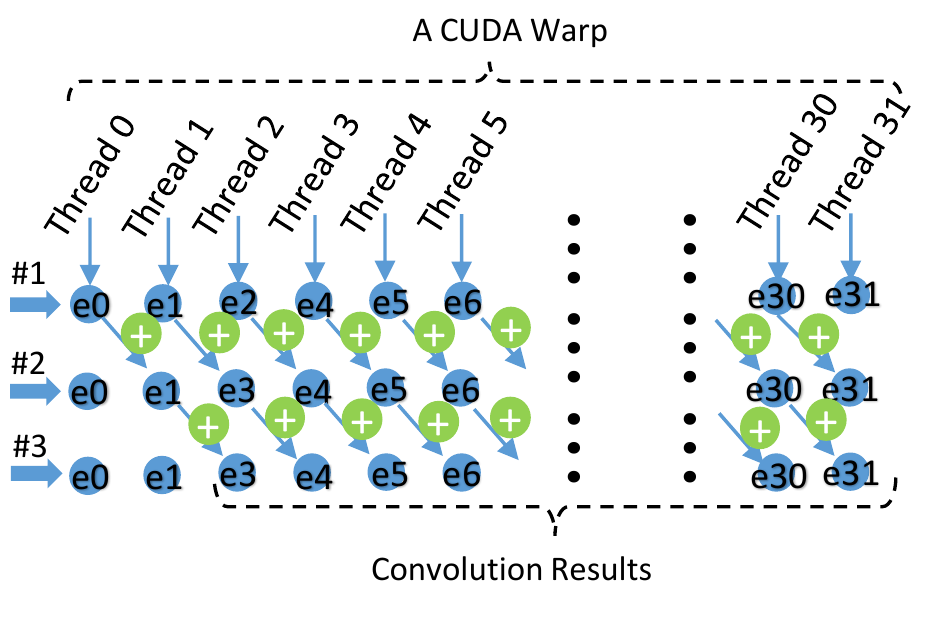}
   \label{shitf_partial_sum}
 }
\caption{(a) Computing 2D convolution by a single warp to 32$\times$C register-cached words and M$\times$N filter matrix. (b) An example of computing partial sums in parallel. (c) Shifting partial results: $e_{i}$ is a partial result computed by thread \emph{i}. Threads (0$\sim$31) represent all threads in a warp. The indicator \#i points to the $i^{th}$ time of shifting partial sums in a warp. $\oplus$ is an addition operator.} 
\label{fig:SSAI}
\end{figure*}

{
Using the SSAM model introduced in the previous section, we implement an efficient 2D convolution algorithm. Regarding the mapping of algorithms to SSAM as \emph{J} (Equation~\ref{equ:model J}), the input \emph{X} and output \emph{Y} are addressed in Section~\ref{sec:regster cache}. The computing operation \emph{O} is discussed in Section~\ref{sec:Parallel Inner Product}, and the dependency graph \emph{D} is discussed in Section~\ref{sec:Shuffle Partial Sum Registers}.
}
The detailed 2D convolution implemented by SSAM model is shown in Listing~\ref{convolution_kerne}:
{
  \setlength{\leftmargini}{15 pt} 
    \begin{enumerate}[label=(\roman*)]
        \item All of the filter weights are stored into shared memory~(lines 7$\sim$12).
        \item A subset of the image data residing in global memory is cached into registers~(lines 13$\sim$14).
    Since registers are a limited resource, the register cache is managed with careful consideration. For this purpose, we introduce a sliding window scheme (more details on that in the next section). 
        \item \label{inner product and shuffle} As shown in Figure~\ref{register_cache_shared_mem_cache}, according to the sliding window\footnote{We use a sliding window to compute several output points per thread. This method improves both data reuse and ILP} position and filter height, we fetch both sub-vector \emph{v} and \emph{w} from the register cache and filter coefficients, respectively. Next, we compute the partial sums by the fused multiply-add operator MAD (lines 24$\sim$26) and transfer the partial sums to the neighbor threads via the \emph{shuffle} primitive (line 22).  
        \item \label{a convolution result} Repeat step \ref{inner product and shuffle} \emph{M} times for all of the sub-vectors ($w_{1}$, $\dots$, $w_{M}$), then store the final partial sums 
    to the register cache again (line 28).
        \item We move the sliding window step by step for a total of \emph{P} times (line 17) in Listing.~\ref{convolution_kerne}. At each step, we repeat the convolution computation, namely \ref{inner product and shuffle} and \ref{a convolution result}.
        \item Finally, the convolution results, which reside in the register cache, are stored back to global memory (lines 30$\sim$31).
    \end{enumerate}
}
The following sections elaborate on the steps of the algorithm.
\lstset{
  language = C++, breaklines = true, breakindent = 10pt, basicstyle = \ttfamily\footnotesize, commentstyle = {\itshape \color[cmyk]{1,0.4,1,0}}, classoffset = 0, keywordstyle = {\bfseries \color[cmyk]{0,1,0,0}}, stringstyle = {\ttfamily \color[rgb]{0,0,1}}, frame = trbl, framesep=5pt, numbers = left, stepnumber = 1, xrightmargin=7pt, xleftmargin=8pt, numberstyle = \scriptsize, tabsize = 3, captionpos = t, directivestyle={\color{black}},  emph={int,char,double,float,unsigned}, emphstyle={\color{blue}},
}
\begin{figure}[t!]
\centering
\begin{minipage}[c]{0.48\textwidth}
\begin{lstlisting}[caption = {SSAM-based 2D Convolution: CUDA kernel. M, N, W, H are defined in Section~\ref{sec:Computing 2D Convolution}, P, B and C are illustrated in Section~\ref{sec:regster cache}. %P and C are defined in Equation~\ref{equ:P} and Equation~\ref{equ:C}~respectively.
}, label = convolution_kerne]
template<typename T, int B, int P, int M, int N>
__global__ void 2Dconvolution(const T* src, 
    T* dst, int W, int H, const T* weight) {
	const int C = P + N - 1;
    //register files
	T data[C];
	__shared__ T smem[N][M];
	T* psmem = &smem[0][0];    
    //1, Load filter weights to shared memory
	for (int i=threadIdx.x; i < N*M; i += B)
		psmem[threadIdx.x] = weight[threadIdx.x];
	__syncthreads();
	//2, Load data from global memory to registers
    data[DATA_IDX] = src[SRC_IDX];
	//3, Compute convolution via registers
	#pragma unroll
	for (int i=0; i<P; i++) {
		T sum = 0;
		#pragma unroll
		for (int m = 0; m < M; m++) {
			if (m > 0)
             sum = __shfl_up_sync(0xffffffff, sum, 1);
			#pragma unroll
			for (int n = 0; n < N; n++) {
			    sum = MAD(data[i + n], smem[n][m], sum);
			}
		}
		data[i] = sum;
	}   
    //4, Store Result to Global Memory
    dst[DST_IDX] = data[DATA_IDX];
}
\end{lstlisting}
\end{minipage}
\end{figure}
\subsection{Register Cache \& Coalesced Memory Access}\label{sec:regster cache}
In SSAM, and in CUDA best practice in general, it is crucial for performance to account for coalesced global memory access. Hence, we make sure that all of the threads in a warp read data from global memory contiguously (one element per thread) as shown in Listing~\ref{convolution_kerne}. The operation is repeated to cache multiple lines of data from global memory into register files, line by line. As illustrated in Figure \ref{register_cache_shared_mem_cache}, each thread in a single warp caches \emph{C} elements as 
\begin{equation}\small \begin{split}
C = N + P - 1
\label{C N P}
\end{split}\end{equation}

Where \emph{N} is the filter size. Each thread computes the output of \emph{P} elements using a sliding window. The sliding window is designed such that a portion of the data in the register cache can be reused when computing the neighboring output points. More specifically, the computation of the convolution of point \emph{P} in a thread can reuse the data in the register cache loaded when computing the convolution of point \emph{P-1}. Using this scheme, at any given point, a WarpSize$\times$C register matrix is stored in the register cache.

In Figure~\ref{register_cache_shared_mem_cache}, the left side of the figure illustrates how to populate the register cache for a warp. In a single warp, each thread reserves C registers for storing data. The register cache size, and systolic array size, in each warp is equal to WarpSize$\times$C. The right side of the figure shows how to cache the filter matrix. We store the filter coefficients in shared memory, then compute the convolution by moving the sliding window step by step \emph{P} times. At each step we compute the inner products of [$v_{i}$, $v_{i+1}$, ..., $v_{i+N-1}$] with $w_{1}$, $w_{2}$, ..., $w_{M}$ as shown in Fig~\ref{parallel_inner_product}. Next, we shift the partial inner product to the neighboring threads as shown in Figure~\ref{shitf_partial_sum}.
{
It is worth emphasizing that such a sliding window provides a simple yet effective method to tackle the second challenge in Section~\ref{sec:introduction}.
}
\subsection{Computing Partial Sums}\label{sec:Parallel Inner Product}
In this section, we describe the computation of partial sums. Computing the partial sum for 2D convolution by Equation~\ref{equ:partial result} is similar to the 1-D convolution example in Section~\ref{sec:1-d-conv}. More specifically as seen in Figure~\ref{parallel_inner_product}, all threads simultaneously, in a systolic fashion, compute the partial sum ($sum_k$) between a register vector \emph{v} ([$v_{i}$, $v_{i+1}$, ..., $v_{i+N-1}$]) and a column of filter $w_{1}$, $w_{2}$, ..., or $w_{M}$. Additionally, the vector v is held by each thread and w is managed by shared memory. It is necessary to access the filter weights in the same order as the data is stored in register cache: namely, unit-strided access in the vertical direction as shown in Figure~\ref{parallel_inner_product}. 
The partial sum requires N multiplications and N-1 addition operations. The multiplication and addition operations are typically optimized to fused-multiply-add (MAD instruction in CUDA~\cite{cudaToolkit}). For the M$\times$N filter, the inner products are computed M times (as shown in Figure~\ref{register_cache_shared_mem_cache}) to compute a single output element of convolution~(Listing~\ref{convolution_kerne} line 20$\sim$27).  
%

\subsection{Transferring \& Accumulating Partial Sums}\label{sec:Shuffle Partial Sum Registers}
In the SSAM model, all of the threads within a single warp accumulate and transfer the partial sums. We use the shuffle instructions to transfer the partial results to the downstream neighbor threads for further accumulations. 
In Figure~\ref{register_cache_shared_mem_cache} the M${\times}$N filter is decomposed into \emph{M} vectors, namely $w_{1}$, $w_{2}$, ..., $w_{M}$. Each partial sum is computed between register vector v~([$v_{i}$, $v_{i+1}$, ..., $v_{i+N-1}$]) and filter vector \emph{w}. Then, all of the inner products, namely partial sums, are shifted to the right side neighbor thread within a single warp using the CUDA \emph{shuffle\_up} function.
In Figure~\ref{shitf_partial_sum}, all of the registers are shifted only once at each step~(Listing~\ref{convolution_kerne} line 22). Next, the shifted partial results are added to the accumulated results by each thread~(Listing~\ref{convolution_kerne} line 25). This process is repeated \emph{M-1} times~(Listing~\ref{convolution_kerne} line 20$\sim$27). Finally a row of convolution results could be attained from a group of threads, namely whose \emph{laneIds}~\cite{cudaToolkit} ranges from \emph{M-1} to \emph{WarpSize-1}. By moving the sliding window once, each warp of threads computes~\emph{WarpSize-M+1} convolution results.    



\begin{figure}[t]
  \centering
  \includegraphics[width=0.32\textwidth]{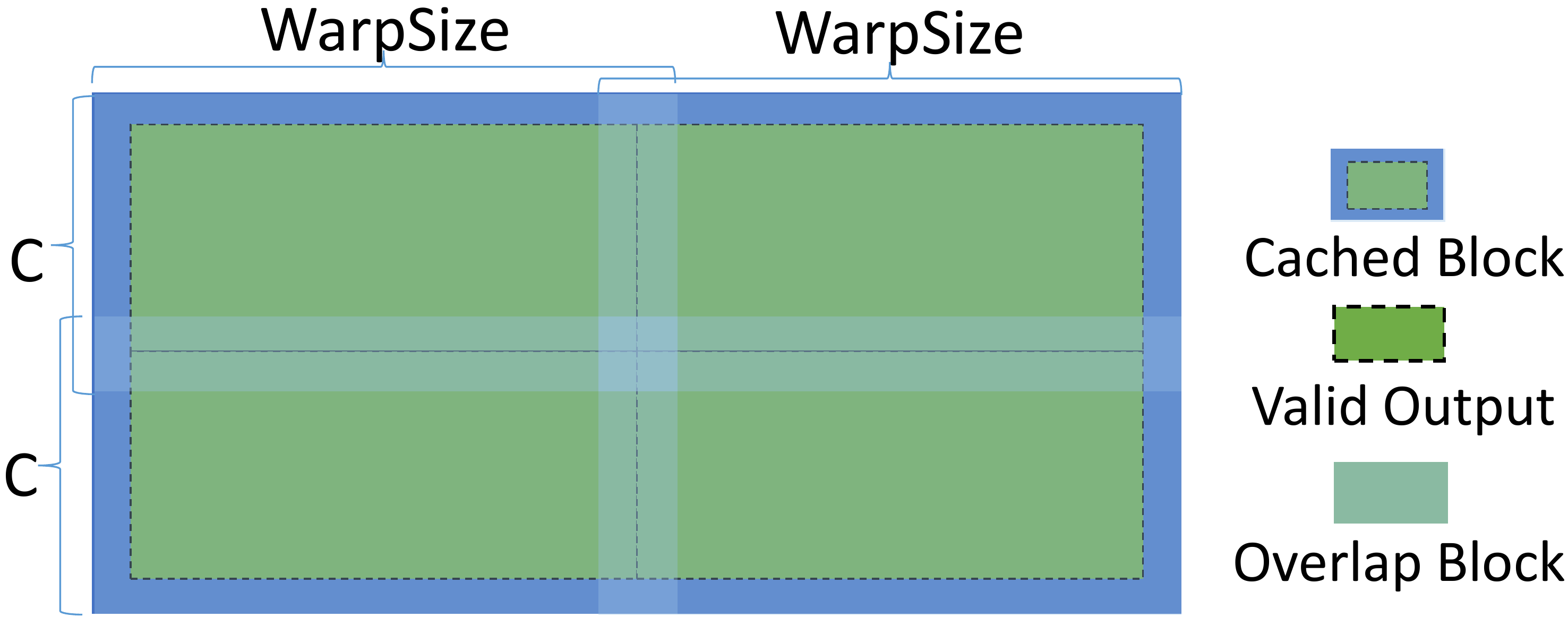}
  \caption{Overlapped Blocking: the cached block size is WarpSize${\times}$C, the valid output size is (WarpSize-M+1)$\times$P. Overlapped block size is a function of the specified filter size.} 
  \label{fig:OverlapBlocking} 
\end{figure}
\subsection{Overlapped Blocking Scheme} \label{sec:overlapped-blocking}
The overlapped blocking scheme is widely adopted to improve the concurrency in time-tiled stencils computations~\cite{DBLP:conf/fpga/ZohouriPM18, Krishnamoorthy:2007:EAP:1273442.1250761}. We use this scheme to eliminate warp-divergence, which can have negative performance effects.
As Figure~\ref{fig:OverlapBlocking} shows, we design our overlapped blocking algorithm to avoid intra-block communications that may cause branching when computing the partial sums and shifting of the register. Using multiple loops for computation is the heaviest and most complex part of computing the 2D convolution (Listing~\ref{convolution_kerne} lines 16$\sim$27).
The overlapped blocking scheme enables all of the threads in our CUDA kernel to perfectly execute without any branching. 
{
Halo layer(s) overhead is introduced when using overlapped blocking (the last challenge in Section~\ref{sec:introduction}). We provide an elaborate analysis of the halo layers impact on performance ( performance model in Section~\ref{sec:halo-layers}).
}

\subsection{Caching Filter Coefficients}
\label{sec:Caching Filter To Shared Memory}
We cache filter weights by shared memory. Filter weights are often tens of bytes (e.g. 36B for 3$\times$3 and \emph{100B} for 5$\times$5). Since the weights are shared by all of the threads in a CUDA block, it is reasonable to access the weights via shared memory. It is also possible to use other kinds of memory, such as constant memory and texture memory. A small number of weights could also be passed to the CUDA kernel directly as arguments. However, considering our commitment to scaling to large filter sizes, we use shared memory in our implementation. 
Our algorithm performs a broadcast read pattern to the shared memory: all threads in a CUDA block access the same address of shared memory. This assures no bank conflict problems, as described by CUDA guide~\cite{cudaToolkit}.

\subsection{Number of Required CUDA Blocks}
We use a one-dimensional block in our implementation, BlockDim=(B, 1, 1).
\emph{B} is equal to blockDim.x, both blockDim.y, and blockDim.z are 1. 
In each CUDA block, its warp count is defined as {\small{$WarpCount = B/WarpSize$}}.
The required CUDA grid dimensions are expressed as 
{\small${GridDim.x=\lceil\frac{W}{WarpCount{\cdot}(WarpSize-M+1)}\rceil}$} and {\small${GridDim.y=\lceil H/P\rceil=\lceil H/(C-N+1)\rceil}$}, respectively.
\begin{figure}[t]
\centering
\begin{minipage}[c]{0.488\textwidth}
\begin{lstlisting}[xleftmargin=5pt, caption = {2D 5-point stencil CUDA kernel. Variable definitions are the same as Listing~\ref{convolution_kerne}. \emph{West}, \emph{North}, \emph{Current}, \emph{South} and \emph{East} are stencil coefficients.
}, label = 5pointStencil kerne]
#pragma unroll
for (int i=0; i<P; i++) {
	T sum = 0;       
    sum = MAD(data[i + 1], West, sum);  //1st column   
    sum = __shfl_up_sync(0xffffffff, sum, 1);
    sum = MAD(data[i + 0], North, sum); //2nd column
    sum = MAD(data[i + 1], Current, sum);
    sum = MAD(data[i + 2], South, sum); 
    sum = __shfl_up_sync(0xffffffff, sum, 1);
    sum = MAD(data[i + 1], East, sum);  //3rd column
    data[i] = sum;
}
\end{lstlisting}
\end{minipage}
\vspace{-10pt}
\end{figure}
\subsection{Mapping 2D Stencil to SSAM}
\label{sec:conv2stencil}
In this section, we describe mapping 2D stencils to SSAM. In Listing~\ref{5pointStencil kerne}, the SSAM-based 5-point stencil kernel differs slightly from the convolution example in Listing~\ref{convolution_kerne}. First, the stencil coefficients are divided into three groups as \{\emph{West}\}, \{\emph{North}, \emph{Current}, \emph{South}\}, and \{\emph{East}\}. Then we compute in parallel the partial sums between the coefficients groups and the cached data. Finally, the partial sums are shifted to the neighbor threads like Section~\ref{sec:Parallel Inner Product}. In 2D convolution, we cache the filter coefficients by shared memory to account for the cases of large filters. Stencils typically have fewer coefficients than convolutions, so we directly transfer the stencil coefficients to the kernel as arguments (namely \emph{Current}, \dots, \emph{East}). 
It is important to note that SSAM is not limited to low order stencils such as the 5-point stencil. SSAM, as will be shown, is highly effective for different shapes of stencils of the high order. 

\subsection{Mapping 3D Stencil to SSAM}
This section discusses using SSAM for a 3D stencil. We divide the 3D grid into many 3D sub-grids with overlapping blocks (to account for the halo layers as in Figure~\ref{fig:OverlapBlocking}). Each sub-grid is processed by a CUDA block, and each warp in a CUDA block processes a 2D slice in the X-Y plane as in Figure~\ref{fig:3dstencil}. Regarding the threads accumulating the final result $s_{i+1}(x, y, z)$, since in the Z direction values of $s_{i}(x, y, z-1)$ and $s_{i}(x, y, z+1)$ are inaccessible (i.e. residing the registers of neighbor warps), we use the shared memory to accumulate the partial sums, which are computed by corresponding warps. In this scenario, SSAM performs intra-warp communication using the shuffle instruction and inter-warp communication using shared memory.

\section{Performance Model} \label{sec:performance-model}
In this section, we introduce a performance model to analyze the effectiveness of SSAM over the conventional scratchpad memory implementations. In addition, we analyze the overhead of processing the halo area required in the overlapped blocking scheme (in Section~\ref{sec:overlapped-blocking}).   
For most of the memory-bound regular applications that could be implemented with SSAM, the analysis results should be valid, yet the details of the analysis may vary from case to case. We use 2D convolution as a motivating example in the following discussions. 

\subsection{Micro-benchmarking} \label{sec:benchmarking}
We use micro-benchmarking to better understand some performance characteristics of GPUs. Several papers have made contributions in demystifying parts of the details about CUDA-enabled GPUs, such as the memory hierarchy latency and throughput~\cite{mei2014benchmarking}. We adopt the micro-benchmarks proposed by cudabmk~\cite{wong2010demystifying}, and improve its functions for our purpose.
Some of the relevant measured results are listed in Table~\ref {tbl:aplt}. 

\begin{table}[t]
\caption{The latency of different operations. All of the latency values are measured by our micro-benchmarks in the unit of cycles/warp. $T_{smem\_read}$ is the latency of reading shared memory.} 
\small
\label{tbl:aplt}
\centering
        \begin{tabular}{?c|cc^c|cc?}
        \bhline
        {\bf{GPU}} & {\bf{Operation}}  &  {\bf{Latency}}  & {\bf{GPU}} & {\bf{Operation}}  &  {\bf{Latency}} \\          
        \bhline
              & shfl\_up\_sync         & 33  &      & shfl\_up\_sync       & 22      \\
        P100  & add, sub, mad          & 6   & V100 & add, sub, mad        &  4      \\ 
              & $T_{smem\_read}$       & 33  &      & $T_{smem\_read}$     &  27     \\               
        \bhline        
        \end{tabular}
\end{table}

\subsection{Efficient In-register Partial Sums Computation}\label{sec:partial-sums}
In this section, we demonstrate the benefits of using SSAM by comparing the latency of computing a single output element of 2D convolution using SSAM versus the conventional shared memory implementation. 

We define the following parameters for modeling the compute time. {\scriptsize$T_{smem\_read}$} is the latency of reading shared memory, {\scriptsize$T_{reg}$} is the latency of reading/writing register, {\scriptsize$T_{shfl}$} is the latency of shuffle instruction, {\scriptsize$T_{mad}$} is the latency of MAD operation, {\scriptsize$T_{gmem\_read}$} and {\scriptsize$T_{gmem\_write}$} are respectively the latency of reading and writing global memory.
Suppose M${\times}$N filter coefficients are cached in shared memory in advance. Conventionally, the image data is cached in shared memory, the latency of computing an output element is
\begin{equation*}\normalsize   
\begin{split}
L_{smem} = M{\cdot}N{\cdot}(T_{mad} + 2{\cdot}T_{smem\_read} + 2{\cdot}T_{reg})
\end{split}
\end{equation*}
Where {\scriptsize$2{\cdot}T_{smem\_read}$} is the time to perform one load of a filter weight plus one store of cached data from shared memory to registers. In the case of SSAM, the image data is cached in register cache, the latency may be expressed as
\begin{equation} \normalsize \begin{split}
L_{reg} = M{\cdot}N{\cdot}(T_{mad} + T_{smem\_read} + 2{\cdot}T_{reg}) + (M-1){\cdot}T_{shfl}
\end{split}\label{equ:L-reg}
\end{equation}
It is worth mentioning that we require (M-1) shuffle instructions for intra-warp communication: no shared memory is required for intra-warp communication. The difference of computing an output element between the shared memory method and register cache can be derived as
\begin{equation}\normalsize  
\begin{split}
    Dif_{smem\_reg} &= L_{smem} - L_{reg} \\
                    &= M{\cdot}N{\cdot}T_{smem\_read} - (M-1){\cdot}T_{shfl}
\end{split}\label{equ:dif-smem-reg}
\end{equation}
Based on the metrics in Table~\ref{tbl:aplt}, The result is {\normalsize$Dif_{smem\_reg}{\gg}0$}, where {$M{\geq}2$} and {$N{\geq}2$}. 

In conclusion, the register-based partial sums computation method is more efficient than the shared memory method.  

\subsection{Halo Layer(s) Overhead}\label{sec:halo-layers}
In this section, we analyze the benefits and overhead of using overlapped blocking (as shown in Section~\ref{sec:overlapped-blocking}) by comparing the shared-memory-cache implementation with SSAM.
The input 2D data residing on global memory is divided into many contiguous blocks as Fig~\ref{fig:OverlapBlocking} shows, with overlapping areas to account for the halo layers. 
Since the cache size is considerably smaller than the original 2D data, the halo data must be stored into cache memory multiple times. Note that the required halo data is loaded from global memory and stored to cache and incurs no additional computation: the halo data does not generate redundancy in the convolution computation.

The following parameters are defined in our analysis: 
$T_{g2rc}$ is the required time of loading data from global memory to register in the case of using the register cache method. When using the shared memory method, $T_{g2rs}$ is the required time for loading the data from global memory to registers and $T_{r2s}$ is required time of storing data from register to shared memory. 

Loading the halo layer data is a redundant operation that could penalize performance. We compare the penalizing effects of halo layers between SSAM and shared memory. We define the ratio of halo for SSAM as
{\tiny $HR_{rc}={(S{\cdot}C-(S-M){\cdot}(C-N))}/{(S{\cdot}C)}$},
where S is WarpSize. Hence, we can further derive that {\tiny $HR_{rc}<{(S{\cdot}N+C{\cdot}M)}/{(S{\cdot}C)}$}.
We define {\scriptsize $HR_{smc}$} as the ratio of shared memory used for the halo layers, where {\scriptsize $HR_{smc}{\in}[0,1)$}, we can reach that {\tiny ${T_{g2rc}}/{(1+HR_{rc})}$} approximates {\tiny ${T_{g2rs}}/{(1+HR_{smc})}$}.
Since the shared memory can be accessed within a CUDA block, while the register can be only within a warp, {\scriptsize $HR_{rc}{\gg}HR_{smc}$} becomes true. Suppose that the shared memory caches the global memory without halo layers as {\scriptsize $HR_{smc}$=0}, we can achieve the ideal caching pattern as {\scriptsize $T_{g2rc} {\approx}(1+HR_{rc}){\cdot}T_{g2rs}$}.
The difference in time between using the shared memory method versus the register cache for loading and storing data may be expressed as 
{\scriptsize $T_{dif\_mem\_io}=(T_{g2rs}+T_{r2s}+T_{r2g})-(T_{g2rc}+T_{r2g})$}, 
where {\scriptsize$T_{g2rc}$} is the time required to store the convolution result of W*H elements back to the global memory. Hence, {\scriptsize$T_{dif\_mem\_io} = T_{r2s} - T_{g2rs}{\cdot}HR_{rc}$}.
Suppose that each thread processes \emph{P} elements and loads \emph{C} elements, the difference of using shared memory and register cache to compute convolution, while accounting for the halo layers is
\begin{equation*}
\scriptsize
\begin{split}
Dif &\approx T_{r2s} - T_{g2rs}{\cdot}HR_{rc}  + P{\cdot}Dif\_{smem\_reg} \\
       &>T_{r2s}-T_{g2rs}{\cdot}({N}/{(N+P-1)}+{M}/{32})+P{\cdot}(M{\cdot}N{\cdot}T_{smem\_read}-(M-1){\cdot}T_{shfl})
\end{split}
\end{equation*}
for each thread loading \emph{C} elements from global memory. The difference of execution time on average can be expressed as
\begin{equation*}
\scriptsize
\begin{split}
AvgDif=Dif/C &> {T_{r2s}}/{C} - {T_{g2rs}}/{C}{\cdot}1/({N}/{(N+P-1)} + {M}/{32}) \\
             & + {P}/{C}{\cdot}(M{\cdot}N{\cdot}T_{smem\_read} - (M-1){\cdot}T_{shfl})
\end{split}
\end{equation*}
In comparison to {\scriptsize$HR_{rc}$}, {\scriptsize$HR_{smc}$} is relatively small since {\scriptsize${T_{r2s}}/{C}$} is purely dominated by the shared memory latency. 
Considering {\scriptsize${T_{g2rs}}/{C}$} as the global memory read latency, we reach {\scriptsize$({T_{r2s}}/{C}) = T_{smem} \gg T_{smem\_read}$} and {\scriptsize$({T_{g2rs}}/{C}){\approx}T_{gmem\_read}$}, thus
\begin{equation*}
\scriptsize
\begin{split}
AvgDif >& T_{smem\_read} - T_{gmem\_read}{\cdot}({N}/{(N+P-1)} + {M}/{32}) \\
        & + {P{\cdot}M{\cdot}N{\cdot}T_{smem\_read}}/{(N+P-1)} - (M-1){\cdot}T_{shfl}
\label{equ:AvgDif}  
\end{split}
\end{equation*}
Since {$T_{gmem\_read}$} is 200$\sim$400 cycles/warp for coalesced access~\cite{nvidia2017compute}.
We can conclude that {$AvgDif{\gg}0$}, where {{M$\geq$2}} and {{N$\geq$2}}. 

\begin{figure*}[t]
\centering
\hspace{0.05cm}
\subfloat[Single precision on P100. cuFFT method is constant as 353 ms.] {
   \includegraphics[width=0.495\textwidth]{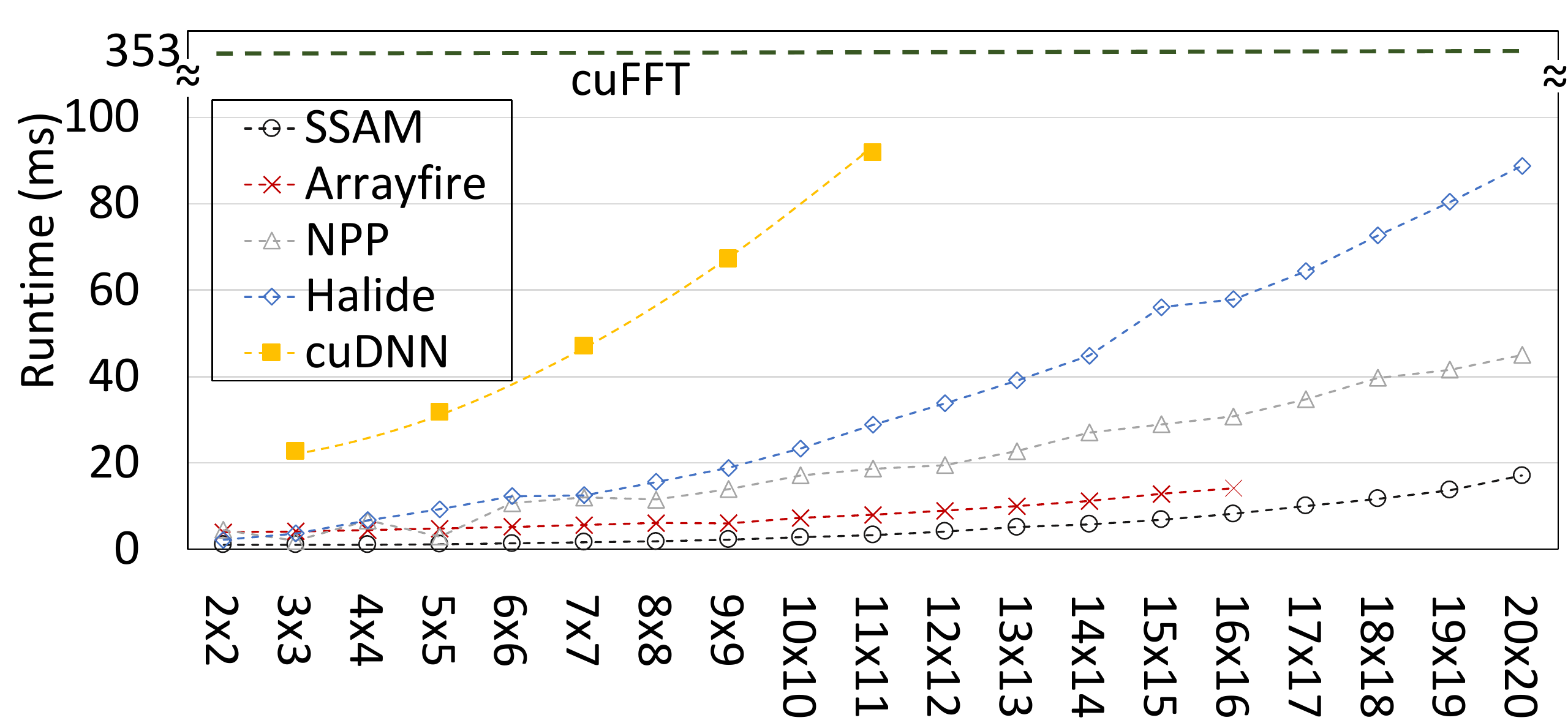}
   \label{fig:2d conv part p100}
 }
\subfloat[Single precision on V100. cuFFT method is constant as 349 ms.] {
   \includegraphics[width=0.495\textwidth]{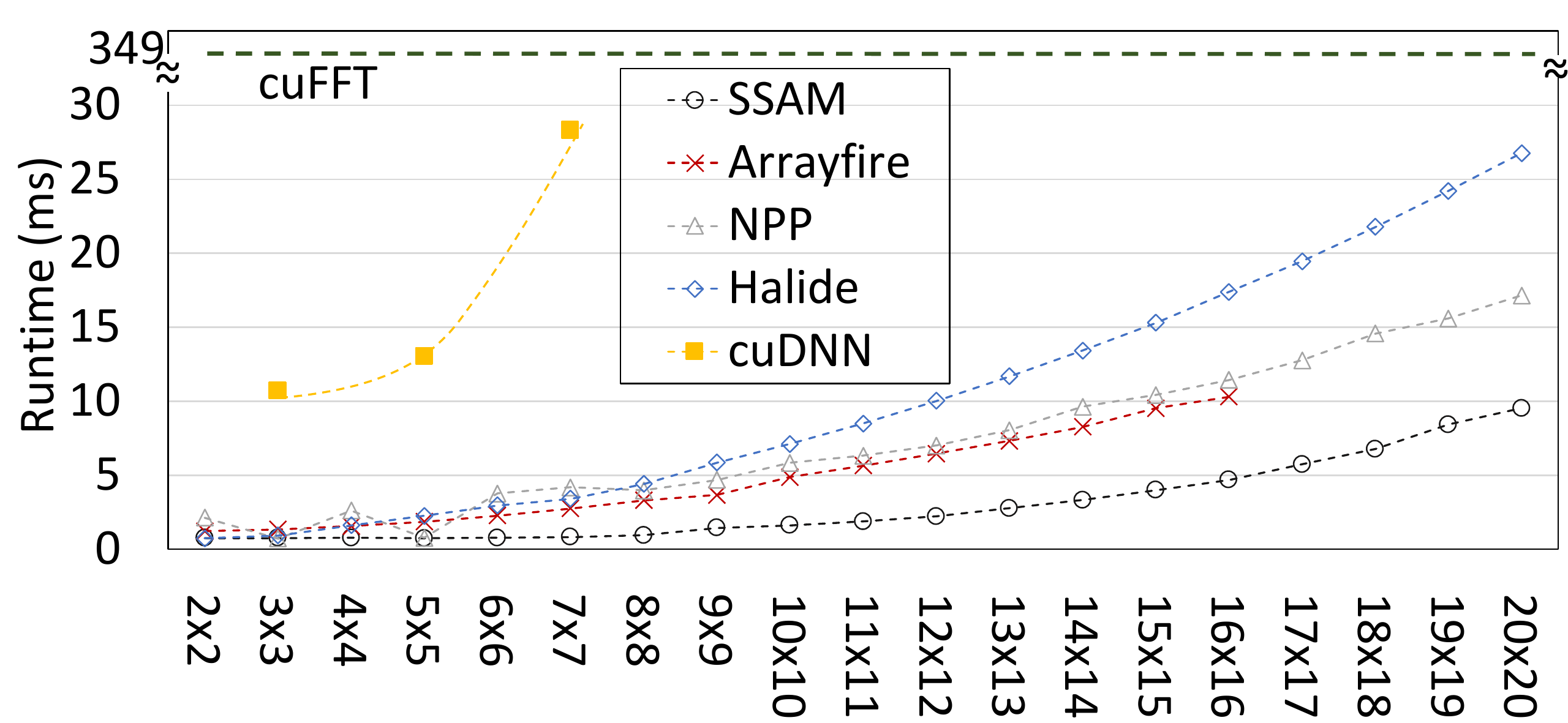}
    \label{fig:2d conv part v100}
 }
\caption{2D Convolution performance and scalability. The image size is $8192{\times}8192$, the x-axis is filter size, the y-axis is execution time.}
\label{fig:2d-conv-p100-v100}
\end{figure*}
To sum up, in comparison to the shared memory-based algorithms, the overhead of handling the halo layers in register cache method is marginal. 
\subsection{The Importance of Dependency \emph{D} in SSAM}
In this section, we present the importance of dependency (namely D) in SSAM. Our performance model proves the bases for why the SSAM improves the performance of memory-bound kernel.  
Based on former discussions and the Equation~\ref{equ:model J}, we can conclude the following:
{
\setlength{\leftmargini}{15 pt}
\begin{enumerate}[label=(\roman*)]
\item The in-register partial sums computing operation (\emph{O}) achieves higher computing efficiency in comparison to the conventional methods.
\item Coalescing access to global memory and efficient register cache make the input (\emph{X}) and the output (\emph{Y}) operations perform data Read/Write efficiently. 
\item However, the partial sums transfer path (\emph{D}) varies from one algorithm to another. Exploring the correct dependency is critical to algorithm performance and thus we need a careful design of the data transfer paths in SSAM. In other words, \emph{D} should be mapped carefully to the register communication pattern within a single warp, e.g. Fig~\ref{fig:ssai} shows that exchanging registers in the horizontal direction is more expensive than vertical. Hence, decreasing the transfer of partial sums in the horizontal direction is essential for expressing algorithms efficiently in SSAM.
{
More specifically, we can determine the best D by computing and comparing the latency of variants of SSAM-based kernels via micro-benchmarking, e.g., computing the latency of 2D convolution as in Equation~\ref{equ:L-reg}. 
}
\end{enumerate}
}

\section{Evaluation} \label{sec:Evaluation}
This section is dedicated to reporting 
the achieved performance of our execution model using the latest Nvidia GPUs.
The performance of 2D convolution and 2D/3D stencil computation is analyzed in detail. It is noteworthy to mention that we prioritized the implementation and reporting of 3D stencil over 3D convolution in the paper, in part due to 3D stencils being more exhaustively studied in literature and targeted with complex data reuse schemes (shown later). We are however currently targeting 3D convolutions after performing an exhaustive study of the diversity of 3D convolutions sizes used in deep learning.

\subsection{Software \& Hardware Setup}
The experimental results presented here are evaluated by two Tesla P100 and V100 GPUs. The CUDA driver is 410.48, we use CUDA 10.0 (including NPP/cuFFT and cuDNN v7.4), GPU memory is 16GB, and the OS is CentOS 7.6.
ECC option for GPUs is on. Nvidia nvcc and gcc-5.4 are used to compile the CUDA kernel and host codes, respectively.

\subsection{2D Convolution Results}
\begin{figure*}[t]
\centering
\subfloat[Single precision on Tesla P100 GPU.] {
   \includegraphics[width=1.0\textwidth]{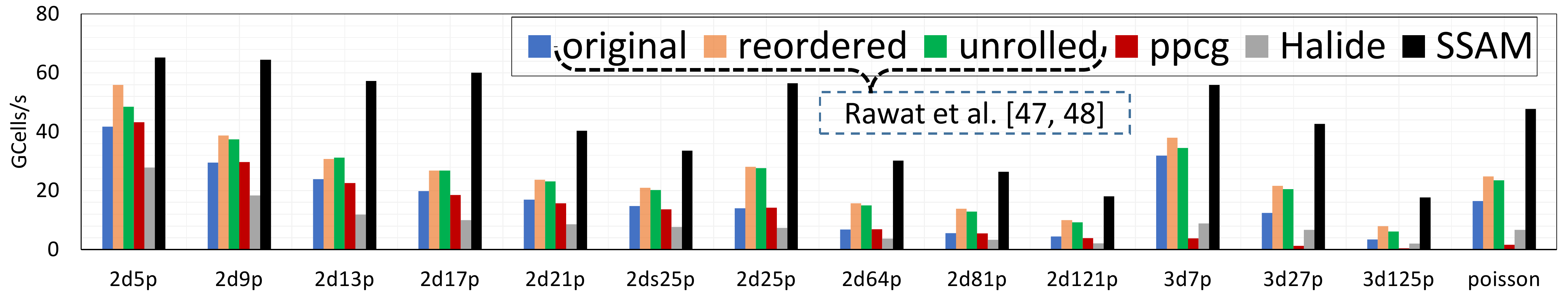}
   \label{fig:stencil-evaluation-p100-float}
 }
 
\subfloat[Single precision on Tesla V100 GPU.] {
   \includegraphics[width=1.0\textwidth]{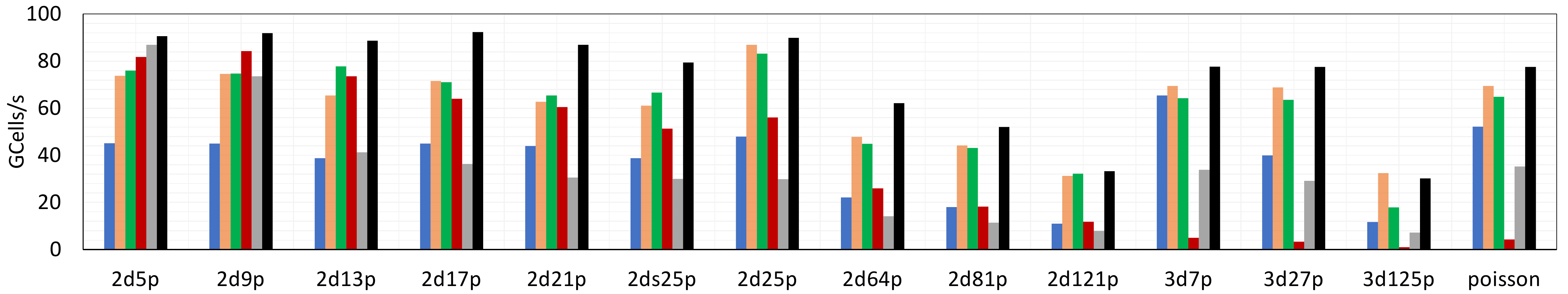}
    \label{fig:stencil-evaluation-v100-float}
 }
 
 \subfloat[Double precision on Tesla P100 GPU.] {
   \includegraphics[width=1.0\textwidth]{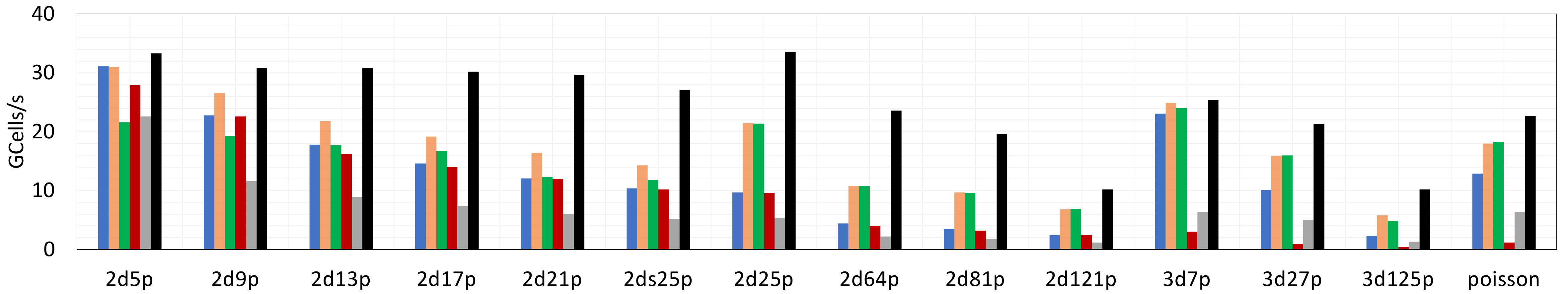}
    \label{fig:stencil-evaluation-p100-double}
 }
 
 \subfloat[Double precision on Tesla V100 GPU.] {
   \includegraphics[width=1.0\textwidth]{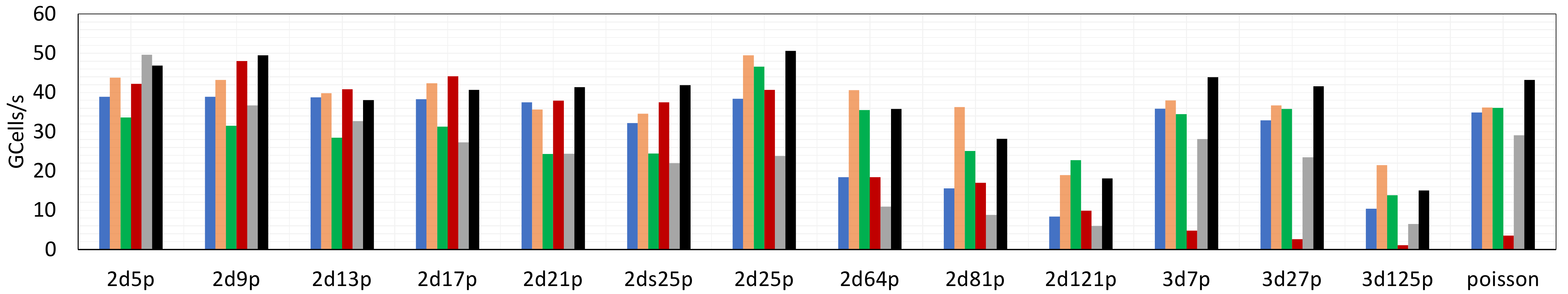}
    \label{fig:stencil-evaluation-v100-double}
 } 
\caption{Performance evaluation for the 2D and 3D stencil benchmarks on Tesla P100 and V100 GPUs, the x-axis is the stencil benchmarks defined in Table~\ref{tbl:stencil-types}. The y-axis is the performance in a unit of GCells/s. The "original", "reordered", and "unrolled" are implemented by Rawat et al. as open-source~\cite{Rawat:2018:ROS:3200691.3178500, Rawat:2018:AIR:3291656.3291718}, the "ppcg" is auto-generated from the C code using the ppcg compiler~\cite{Verdoolaege2013PPCG}.}
\label{fig:stencil-evaluation}
\end{figure*}

In Figure~\ref{fig:2d-conv-p100-v100}, we evaluate 2D convolutions for various filter sizes, which ranges from 2$\times$2 to 20$\times$20.
It is noteworthy that P=4, B=128 are used as parameters. We use the OS-independent \emph{cudaEvent} function to measure the execution time for all of the convolution computations, and ignore the data transfer time between host and device. Note that although in our experiments we only show square-shaped filters (i.e. M=N), a simple change in a template function in our implementation enables the computation of 2D convolution for any filter shape (M$\neq$N) as well. 

{
\setlength{\leftmargini}{14 pt}
\begin{enumerate}[label=(\roman*)]
\item {\bf{ArrayFire}}~\cite{Yalamanchili2015}.~
ArrayFire is a highly optimized library by CUDA. We report the performance of the fastest 2D convolution kernel, called \emph{kernel::convolve2}, in which the shared memory and constant memory are employed to cache image data and filter weights, respectively.  
Its filter size limitation is 16$\times$16. Note that there are no official documents about this value, which was found out by analyzing source code and lots of tests. 
\item {\bf{NPP}}~\cite{cudaToolkit}.~
The Nvidia Performance Primitives (NPP) library is a closed-source library, thus we investigate it by the \emph{nvprof} profiling tool. NPP does not use any shared memory as cache. Note that NPP particularly optimizes the convolution computation for filters of size 3${\times}$3 and 5${\times}$5 using dedicated kernels, the kernel names are \emph{FilterBorder32f3x3ReplicateQuadNew} and \emph{FilterBorder32f5x5ReplicateQuadNew}, respectively.

\item {\bf{cuFFT}}~\cite{cudaToolkit}.~
The Nvidia CUDA Fast Fourier Transform (cuFFT) library provides a constant, yet relatively high, run time regardless of the filter size.

\item {\bf{Halide}}~\cite{mullapudi2016automatically}.~
The Halide is a domain-specific language (DSL) designed to 
generates pipelines for image processing~\cite{mullapudi2016automatically}. It is difficult to instrument its pipeline to measure the kernel execution time, so we use the Nvidia \emph{nvprof} profiling tool to report the kernel execution time.
\item {\bf{cuDNN}}~\cite{chetlur2014cudnn}.~
cuDNN is a GPU-accelerated library by Nvidia for deep neural networks.
We evaluate it by a special parameter that an image with a single channel is convolved by a single filter. Several algorithms are implemented in cuDNN, and we only report the result with the best performance. Note that the supported filter sizes in cuDNN must be odd numbers, e.g. $3{\times}3$, $5{\times}5$.

\end{enumerate}
}

Figure~\ref{fig:2d-conv-p100-v100} gives indications of the accuracy of the findings of the performance model (Section~\ref{sec:performance-model}). Equation~\ref{equ:dif-smem-reg} indicates that given a fixed output size of 2D convolution, while increasing size of filters, the performance difference between SSAM and other kinds of implementations (not only shared memory-based algorithm) becomes increasingly larger. Such a conclusion can be verified by observing the difference in the performance of the graphs in Figure~\ref{fig:2d-conv-p100-v100}. 


    \begin{table}[t] 
          \footnotesize
          \caption{Stencil benchmark. The k is stencil order, the FPP is FLOP per point. The domain sizes of the 2D and 3D stencil are $8192^2$ and $512^3$, respectively. 
          A detailed description of the benchmarks can be found in literatures~\cite{Rawat:2018:ROS:3200691.3178500,Rawat:2018:AIR:3291656.3291718}.
          }
          \centering
          \begin{tabular}{ | c  c  c  ?  c  c  c ? c  c  c | }
          \bhline
          \bf{benchmark}        & \bf{k}& \bf{FPP} & \bf{benchmark}       & \bf{k}& \bf{FPP} &\bf{benchmark}         & \bf{k}& \bf{FPP}\\
          \bhline
           2d5pt          & 1     & 9        & 2d9pt          & 2     & 17        & 2d13pt         & 3     & 25\\
           2d17pt         & 4     & 33       & 2d21pt         & 5     & 41        & 2ds25pt        & 6     & 49\\ 
           2d25pt         & 2     & 33       & 2d64pt         & 4     & 73        & 2d81pt         & 4     & 95\\ 
           2d121pt        & 5     & 241      & 3d7pt          & 1     & 13        & 3d13pt         & 2     & 25\\ 
           3d27pt         & 1     & 30       & 3d125pt        & 2     & 130       & poisson        & 1     & 21\\
          \bhline
          \end{tabular}  
          \label{tbl:stencil-types} 
      \end{table}        
\subsection{Stencil Results}
To better understand the performance of SSAM-based stencil computations, we evaluate a diverse collection of 2D/3D stencil benchmark (listed in Table~\ref{tbl:stencil-types}) and compare the performance with a variety of CUDA implementations on the latest GPUs.
Note that the stencils in the experiments include both low order and high order stencils.
High-performance stencil libraries rarely provide a consistent performance advantage for both low and high order stencils since high order stencils are typically bound by the registers. For instance, temporal blocking is effective for low order stencils~\cite{DBLP:conf/fpga/ZohouriPM18}, while register re-ordering schemes are effective for high order stencils~\cite{Rawat:2018:ROS:3200691.3178500, Rawat:2018:AIR:3291656.3291718}. 

Figure~\ref{fig:stencil-evaluation} shows three implementations: "\emph{original}", "\emph{reordered}", and "\emph{unrolled}", with state-of-the-art performance results by Rawat et al.~\cite{Rawat:2018:ROS:3200691.3178500, Rawat:2018:AIR:3291656.3291718}. The "\emph{original}" is a basic CUDA implementation," \emph{reordered}" is register-optimized version for "\emph{original}", "\emph{unroll}" is the unroll-optimized method. Note that we adapted Rawat's source code to P100/V100 GPUs, ran multiple tuned configurations and only report the result with the best performance. We also compare with the latest \emph{ppcg} (0.08 version) compiler~\cite{Verdoolaege2013PPCG} and Halide~\cite{mullapudi2016automatically}. 

We use GCells/s and not GFlops/s as the performance metric in all our stencil experiments since we observed that different libraries count FLOPS differently. Additionally, the achieved GFlops/s can be obtained by multiplying the reported GCells/s with the corresponding FPP factor (as in Table~\ref{tbl:stencil-types}). Note that the execution time of all stencil kernels is measured by Nvidia \emph{nvprof} profiler.



In Figure~\ref{fig:stencil-evaluation}, SSAM mostly outperforms the other implementations using the latest GPUs in both single and double precision runs. The performance advantage of SSAM is due to the better register locality, less global memory access and efficient threads communication. Such a result is consistent with the performance of 2D convolution (as in Figure~\ref{fig:2d-conv-p100-v100}) and is further explained by the performance model. 

There are two important observations in Figure~\ref{fig:stencil-evaluation}. One is that in comparison to the P100 GPU, the performance variance on V100 become smaller, and the other is that when using double precision on the V100 GPU, few of the SSAM-based stencils do not achieve the highest performance. The reason behind that is that Volta's architecture is different in many aspects, which will be discussed in Sec~\ref{sec:discussion_p100v100}. 

\begin{figure}[t]
  \centering
  \includegraphics[width=0.497\textwidth]{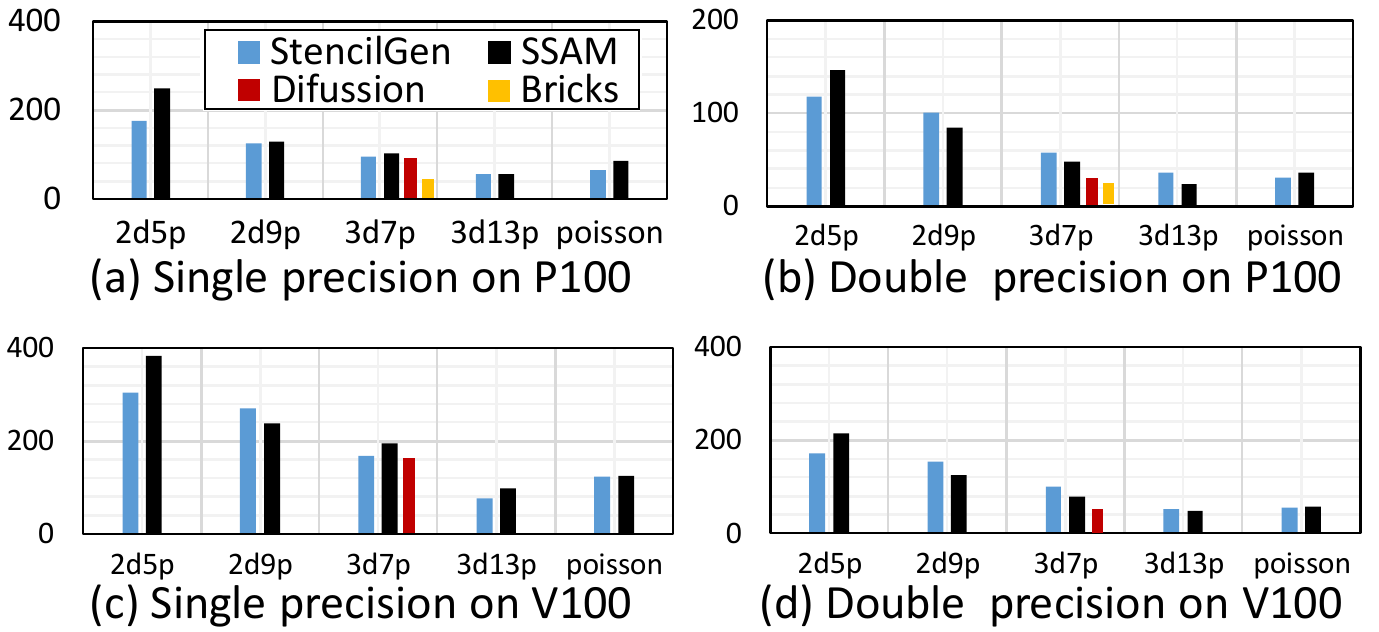}
  \caption{Performance evaluation for the 2D and 3D stencil computation using temporal blocking. the x-axis is stencil benchmark, the y-axis is the performance in the unit of GCells/s.} 
  \label{fig:stencil-temporal_blocking} 
\end{figure}
\subsection{Comparison with Temporal/Spatial Blocking Libraries}
SSAM is a versatile model that enables temporal blocking without much change to the implementation: the use of register cache and shuffling in SSAM does not limit the use of temporal blocking. 
It is worth to mention that temporal/spatial blocking algorithms are widely used to improve the performance of stencil computation by reducing the required memory bandwidth.
In this section, we focus on comparing SSAM to state-of-the-art temporal blocking libraries for stencils. 

\paragraph{\bf{StencilGen}}We compare the performance with {\bf{StencilGen}}~\cite{Rawat-stencilGen}.
To the author's knowledge, StencilGen reports the highest performance among temporal blocking libraries.
{
As Figure~\ref{fig:stencil-temporal_blocking} shows, in the majority of benchmarks, SSAM performs better than StencilGen. However, register pressure can limit our performance in some cases.
It is important to mention that StencilGen is optimized for stencil computations only, while SSAM is more versatile and general for different types of kernels.}
In addition, it is important to point out that SSAM provides a consistent performance advantage for both low and high order stencils. Some approaches, such as those proposed in~\cite{Rawat:2018:AIR:3291656.3291718, Rawat-stencilGen}, can further be adopted by SSAM to improve the performance for the specific cases of deep temporal blocking.


\paragraph{\bf{Diffusion}}The authors in~\cite{DBLP:conf/fpga/ZohouriPM18} report a competitive performance of the 3d7pt stencil (highly optimized by shared memory as proposed in~\cite{maruyama2014optimizing}, called {\bf{Diffusion}} in Figure~\ref{fig:stencil-temporal_blocking}) as 92.7~GCells/s and 162.4~GCells/s using single precision on P100 and V100 GPUs, 30.6~GCells/s and 46.9~GCells/s using double precision on P100 and V100 GPUs, respectively. 
{
This performance is significantly lower than SSAM.}

\paragraph{\bf{Bricks}} {\bf{Bricks}} is a general stencil library targeting both CPUs and GPUs~\cite{zhaodelivering}. The performance of Bricks on P100 is reported to be 41.4 GCells/s and 24.25 GCells/s for single and double precision, respectively. As seen in Figure~\ref{fig:stencil-temporal_blocking}a and Figure~\ref{fig:stencil-temporal_blocking}b, SSAM outperforms Bricks on P100. Since Bricks is not publicly available, we can not compare our result for the V100 GPU.

{
\section{Discussion}\label{sec:discussion}
This section discusses the performance considerations given to Pascal and Volta architectures, and elaborates on how to extract dependency for SSAM.
\subsection{\bf{Considerations for Pascal \& Volta Architectures}}\label{sec:discussion_p100v100} 
In this section, we outline the reasons why SSAM performs differently on P100 and V100 GPUs.
}
{
    \setlength{\leftmargini}{12 pt}
    \begin{enumerate}[label=(\roman*)]
    \item 
    The L1 cache in Volta is significantly enhanced. 
    (1) According to Nvidia user guide~\cite{tuning2018cuda}, the {\bf{capacity}} of L1 cache in Volta has increased up to 128KB, which is more than 7$\times$ larger than Pascal; 
    (2) Regarding the access {\bf{latency}} of L1 cache, V100 is about 2.8$\times$ faster than the P100 (as reported in~\cite{jia2019dissecting}: V100 is 28 cycles vs. P100 is 82 cycles).
    This narrows the gap between applications that are manually optimized by caching data in registers (or shared memory) and those that access global memory directly. Hence, the difference of performance between the SSAM and other implementations become smaller on the V100 GPU.
    
    \item{
    The improvement of L2 cache also contributes to the change in performance behavior between P100 and V100 GPUs.
    According to the experiments in~\cite{jia2019dissecting}, in comparison to P100, the {\bf{capacity}} of L2 cache in V100 increases by 50\% (P100 is 4096KB vs. V100 is 6144KB), the access {\bf{latency}} improves by up to 10\% (P100 is $\leq$234 cycles vs. V100 is $\leq$193 cycles).}
    

   
    \item As Jia et al.~\cite{jia2018dissecting} illustrated, due to register bank conflict in Volta architecture, MAD instruction (accessing 3 registers in the same cycle) results in one clock stall. More specifically, in Volta, registers are divided into two banks. However, other generations (e.g. Pascal, Maxwell, Kepler) have four banks~\cite{anderson2012predictive,zhang2017understanding}. Finally, SSAM performs as well as Pascal in Maxwell and Kepler architectures. Due to the space limitation, we do not show the result. 
    \end{enumerate}
}
{
\subsection{\bf{Automation and Extracting Dependency (D) for SSAM}}\label{sec:automation-ssam}
This paper proposes a method for executing kernels with regular data-access behavior in a systolic fashion. The paper is dedicated to explaining the method in a comprehensive way that covers the systolic representation/mapping, a performance model, detailed implementation, and a broad range of experiments and comparisons. 
It is important to note that all kernels evaluated in Section~\ref{sec:Evaluation} are strictly implemented as Listing~\ref{convolution_kerne} and Listing~\ref{5pointStencil kerne} without extra manual optimization. Taking Figure~\ref{fig:2d-conv-p100-v100} for instance, Listing~\ref{convolution_kerne}, a single template kernel function, is used for all SSAM-based 2D convolution evaluations. 

That being said, it is a non-trivial task for users to manually apply SSAM in large code bases made of tens kernels. An ideal case would be to automate SSAM, by using a DSL or code transformation. We argue that code transformation is the more practical approach since users can be reluctant to move to new DSLs. To apply SSAM as an automated code transformation, extracting the dependency and mapping it to SSAM is an important point to consider when generalizing SSAM to different types of algorithms. 
Automatic code generation by SSAM is future work that is outside the scope of this paper due to the space limit, and the nontrivial body of work required to extract dependencies and map them to SSAM for any given kernel. 

In this paragraph, we briefly discuss the possibility of representing the dependency graph (namely the D in Equation~\ref{equ:model J}) using the polyhedral model~\cite{WILDE1993POLYLIB,Griebl98codegeneration,LOECHNER1999POLYLIB}. The polyhedral model is a well-researched mathematical framework for performance optimization~\cite{benabderrahmane2010polyhedral}, which often involves nested loops and large numbers of operations, i.e. convolution, matrix multiplication. According to our formulation in Equation~\ref{equ:model J}, and motivational examples in earlier sections, such a parametric representation can be analyzed for mapping algorithms and code, i.e. C/C++, to the SSAM-based intermediate representation (IR). 
    
}
\section{Related Work}
\label{sec:Related Work} 
{\emph{Register cache}} methods are widely used on GPUs to boost individual application's performance~\cite{lai2013performance,enfedaque2015implementation,hou2017fast,ben2016fast}. 
Most notably, a technology called register packing demonstrated how to avoid shared memory communication for improving cyclic reduction performance~\cite{davidson2011register}.
Lai and Seznec~\cite{lai2013performance} suggest a method named register blocking to explore the potential peak performance of SGEMM on Fermi and Kepler GPUs.
Enfedaque et al. proposed a discrete wavelet transform (DWT) implementation for fast image processing based on register cache~\cite{enfedaque2015implementation}. 
The binary finite field multiplication algorithm was implemented by Eli Ben-Sasson et al. yielded up to 138$\times$ speedup than the popular Number Theory Library~\cite{ben2016fast}.
Hou et al.~\cite{hou2017fast} implemented a register-based sort method shows great improvements over scratchpad memory methods on NVIDIA K80-Kepler and TitanX-Pascal GPUs.
A 1-D stencil method is introduced as an example to illustrate how register cache and shuffle instruction works~\cite{ben2016fast}. 

{\emph{Register optimization}} is an important approach to improve the performance of stencil codes. 
The authors in \cite{zhangmodeling} implemented a register-only method to improve the 3D 7-point stencil without using thread communication. 
Zumbusch et al. implemented vectorized kernels for high order finite stencils on multi-platforms, i.e. AMD/Intel CPUs, Nvidia GPUs~\cite{zumbusch2012vectorized}.
Rawat et al. proposed a reorder framework to optimize register allocation for both CPUs and GPUs~\cite{Rawat:2018:ROS:3200691.3178500, Rawat:2018:AIR:3291656.3291718}.

\emph{Partial sums} method is used by Basu et al. to generate SIMD code for CPUs within CHiLL compiler~\cite{basu2015compiler}. On the other hand, SSAM is introduced at the application level to perform the in-register computation and not just partial sums for GPUs.
Krishnamoorthy et al. used \emph{overlapped tiling} optimization for the stencil to preserve concurrency in time-tiled computations~\cite{Krishnamoorthy:2007:EAP:1273442.1250761}. 

The majority of the work above is limited in scope to specific applications (e.g. stencil), and architectures. Moreover, they lack performance models for the redundancy and do not optimize for reuse in register cache. 
Li et al. proposed a general model for all generations of GPUs, called Warp-Consolidation, however, it is focused on improving the Cooperative-Thread-Array execution~\cite{li2018warp}. 

{
StencilGen~\cite{Rawat-stencilGen}, a state-of-the-art DSL implementing advanced temporal blocking for stencils. Halide~\cite{mullapudi2016automatically} is a DSL and ppcg~\cite{Verdoolaege2013PPCG} is general code-to-code transformation framework. Both generate code for GPUs, yet the automated code does not always yield the ideal performance as demonstrated in earlier sections.
}

{
Using a Stateful DataFlow multiGraph (called as SDFG) to optimize HPC applications, Ben-Nun et al.~\cite{ben2019stateful} demonstrated a very general IR (Intermediate Representation) for a number of accelerators, e.g. CPUs, GPUs, FPGAs. Obviously, SDFG is reported to achieve high performance by considering more characteristics of GPU architectures.
}

To the best of our knowledge, this work is the first to propose a completely generic and versatile method, driven by a performance model, to use locality-optimized register cache and shuffle instruction for directly computing different filter/stencil sizes for both 2D and 3D grids.
{
Last but most importantly, we pave a novel way to optimize algorithms on CUDA-enabled GPUs : mapping the CUDA cores to software systolic arrays. Thereby, we build a bridge between code automation and optimization.
}

\section{Conclusion and Future work}\label{sec:Conclusion} 
This paper proposes a novel and versatile high-performance execution model (namely SSAM) for improving the performance of structured grid memory-bound kernels on CUDA-enabled GPUs. Using the SSAM execution model, we improve the performance of challenging problems, such as convolution and stencil. Our model performs as a software systolic array by computing and shifting partial sums. The evaluation shows that SSAM outperforms top libraries for 2D convolution and 2D/3D stencil, in both Pascal and Volta GPU architectures. 

{
Nvidia Volta V100 GPU includes hardware systolic arraya (namely Tensor cores). Tensor cores are efficient in computation yet they are limited in precision since they designed for Deep Learning workloads. Our software systolic array can be used for a varity of applications at single and double precision.}
For future work, we plan to apply our model to 3D/4D convolution workload for accelerating deep learning training.
We also intend to generate SSAM codes automatically using polyhedral analysis and DSL tools, e.g. ppcg, Halide.

\section*{Acknowledgment}
This work was partially supported by JST-CREST under Grant Number JPMJCR1687. 
Computational resource of AI Bridging Cloud Infrastructure (ABCI) provided by National Institute of Advanced Industrial Science and Technology (AIST) was used. 
A part of the numerical calculations were carried out on the TSUBAME3.0 supercomputer at Tokyo Institute of Technology.
We would like to thank Endo Lab at Tokyo Institute of Technology for providing a Tesla V100 GPU Server.